\documentstyle[aps,prb,psfig,twocolumn]{revtex}
\renewcommand{\min}{\mathop{\rm min}\nolimits}

\renewcommand{\min}{\mathop{\rm min}\nolimits}

\def\lapprox{\,\raise0.4ex\hbox{$<$}\kern-0.8em\lower0.7ex\hbox{$\sim$}\,}
\def\gapprox{\,\raise0.4ex\hbox{$>$}\kern-0.8em\lower0.7ex\hbox{$\sim$}\,}
\begin{document}
\bibliographystyle{prsty}
\title{ \begin{flushleft}
{\small
\sc
PHYSICAL REVIEW B
\hfill
Volume 65, Number 19-20
\hfill
 15 May 2002} \\
\vspace{2mm}
\end{flushleft}
 Activation  energy in a quantum Hall ferromagnet
  and non-Hartree-Fock Skyrmions} 
\vspace{-5mm}
\author{
S. Dickmann${}\!\!$}
\address{ Max Planck Institute for Physics of Complex Systems 
N\"othnitzer Str. 38, D-01187 Dresden, Germany}
\address{Institute for Solid State Physics of Russian Academy of Sciences,
142432 Chernogolovka, Moscow District, Russia\\
\smallskip
{\rm(Received 3 August 2001; revised manuscript 17 December 2001)}
\bigskip\\
\parbox{14.2cm}
{\rm
The energy of {\it Skyrmions} is calculated with the help of a
technique based on the
excitonic representation: the basic set of one-exciton states is used for
the perturbation-theory formalism instead of the basic set of one-particle
states.
We use the approach, at which a skyrmion-type excitation (at zero
Land\'e factor)
is considered as a smooth non-uniform rotation in the 3D spin space. The result
within the framework of an excitonically diagonalized part of the Coulomb
Hamiltonian can be obtained by any ratio $
  r_{\mbox{\tiny C}}=(e^2/\varepsilon {}l_B)/\hbar \omega_c
$ [where $e^2/\varepsilon {}l_B$ is the typical Coulomb energy
(${}l_B$ being the magnetic length); $\omega_c$ is the cyclotron
frequency], and the Landau-level mixing is thereby taken into
account. In parallel with this, the result is also found exactly,
to second order in terms of the $r_{\mbox{\tiny C}}$ (if supposing
$r_{\mbox{\tiny C}}$ to be small) with use of the total
Hamiltonian. When extrapolated to the region $r_{\mbox{\tiny
C}}\sim 1$, our calculations show that the skyrmion gap becomes
substantially reduced in comparison with the Hartree-Fock
calculations. This fact brings the theory essentially closer to
the available experimental data.
\begin{flushleft}
\end{flushleft}
}
}
\maketitle

\centerline{\bf I. INTRODUCTION}

$\;$

Up to now, in two-dimensional (2D) electron gas (EG) the quantum
Hall effect (QHE) at the Landau-level (LL) filling factor of
$\nu=1$, has attracted much
theoretical\cite{so93,mo95,kr95,fe94,by96,fe97,io99,io00} and
experimental\cite{ba95,ku97,ba96,sc95,ma96,le98,me99,sh00}
attention. The interest is explained by the fact that in this
regime at vanishing (or considerably reduced) Zeeman coupling
peculiar Fermionic excitations exist: namely, these are {\it
skyrmions}, characterized by a large spin, $|\delta S_z|\gg 1$,
and by a topological invariant (or topological charge) in terms of
a field theory of the classical 2D ferromagnet.\cite{be75,ra89}
The first mapping of the spin-polarized quantum Hall system to an
appropriate nonlinear $O(3)$ model, of the 2D ferromagnet, was
used in the work of Sondhi {\it et al}$\;$  \cite{so93} to
calculate the skyrmion energy. Within the context of the 
phenomenological approach employed
 the creation gap (i.e, the combined
energy of a skyrmion-antiskyrmion pair) was found in the ideal 2D
case to be exactly equal to half the gap in creating an
electron-hole pair. (The latter is considered as an extreme case
of a spin exciton with $\delta S_z=-1$.) Due to this result, at
once the theory became by factor of 2 closer to the data found
experimentally for the QHE activation gap. However, this and all
later calculations, still remain in striking discrepancy with
measurements. There is growing experimental evidence
\cite{ma96,le98,me99,sh00} that at zero Land\'e factor, the energy
required for activation of a dissipative current in a quantum Hall
ferromagnet, is approximately a factor of 0.1 smaller than the
calculated skyrmion-activation energy (the latter is one-half the
creation gap).

Justification for the application of the nonlinear $O(3)$
model to the quantum Hall ferromagnet (QHF) is confirmed by a
microscopical theory based on the Hartree-Fock (HF) Hamiltonian
for the Coulomb interaction and on the approximation of wave
functions projected (WFP) onto a single LL.\cite{fe94,by96,fe97}
In these works, the energy of isolated skyrmionic excitations is
recalculated and  the minimum creation gap becomes the same as in
Ref. \onlinecite{so93}. Another approach, developed by Iordanskii
{\it et al}$\;$ \cite{io99} does not use the approximation of WFP.
The authors describe a skyrmion excitation as a smooth nonuniform
rotation in real three-dimensional spin space. Due to the fact
that the Coulomb Hamiltonian is invariant with respect to such a
rotation, this approach has an evident advantage. The authors
calculate the energy of skyrmionic excitations with the help of
the perturbation theory technique. This theory uses for a bare
Green function (GF) the appropriate mean-field one-electron GF. In
doing so, the HF approximation is employed and the results (after
a small correction\cite{io00}) turn out to be in agreement with
earlier results.\cite{fe94,by96} (See below in the Appendix II.)
Naturally, any effects of LL mixing are neglected there, and  the
results reported in Refs. \onlinecite{fe94,by96,fe97,io99,io00},
as well as in Ref. \onlinecite{so93},  represent  the energy of
skyrmions only in terms of a linear approximation of the Coulomb
interaction; i.e. only in the framework of the first-order
approximation in the parameter $
  r_{\mbox{\tiny C}}
$
which is supposed to be small.

In our work the energy of skyrmions is calculated analytically
with the help of the modified perturbation-theory technique. This
technique is based on the excitonic representation (ER) which is
suitable when a 2D EG is in a dielectric state, i.e. in the
absence of free electrons and holes (see Refs.
\onlinecite{dz83-84,dz91,by87,di96,dile99,di99,diE99,di00}). {\it
Neither the  HF nor WFP approximations are used}. As in the case
of Ref. \onlinecite{io99}, a skyrmion excitation is considered as
a rotation in 3D space:
$$
  {\vec \psi}({\bf r})={\hat U}({\bf r}){\vec \chi}({\bf r}),\quad
 {\bf r}=(x,y)\,.           \eqno (1.1)
$$
Here ${\vec \psi}$ is a spinor given in the stationary coordinate system
and ${\vec \chi}$ is a new spinor in the local coordinate system
accompanying this rotation. The rotation matrix of ${\hat U}({\bf r})\quad$
($U^{\dag}U=1$) is
parametrized by three Eulerian angles.\cite{ll91} In the zero approximation
in terms of ${\hat U}({\bf r})$ gradients\cite{foot1}, we get
${\vec \chi}\propto {1\choose 0}$.
Generally, in the limit of a vanishing Land\'e
$g$ factor, the results obtained  may be presented as an exact expansion in
terms of the parameter $r_{\mbox{\tiny C}}$. Virtually {\it two first terms of this
expansion have been calculated exactly}.

As a first step, we ignore the Coulomb-interaction processes
responsible for any decay of a one-exciton state due to the
transformation into two-exciton states. Specifically, the 
part of the Coulomb Hamiltonian kept involves:  first, the terms
responsible for the direct Coulomb interaction without any LL
mixing; second, the terms providing the shift of the exchange
self-energy if an electron is transferred from one LL to another;
and finally, the random-phase-approximation terms in which an
exciton transforms into another one at a different point of
conjugate space, but with the same spin state and the same
cyclotron part of energy. In other words, such a reduced
Hamiltomian corresponds to a proper mean-field approach formulated
in relation to excitons, but not to a quasiparticle excitation (as
it would be in the case of the HF approximation). This Hamiltonian
is ``excitonically diagonalized'' (ED)  and  all one-exciton
states (spin waves, magnetoplasmons and spin-flip magnetoplasmons)
present a full basic set for its exact diagonalization.\cite{ka84}
The {\it main idea} of the present work is to employ a basic set
of one-exciton states for the perturbation-theory formalism, instead
of the basic set of one-particle states. If we restrict our study
to the terms of the excitonically diagonalized Hamiltonian (EDH),
then the energies of skyrmion excitations may be found in a very
simple way {\it for any magnitude of} $ r_{\mbox{\tiny C}}$. In
the QHF, at $\nu=1$, and in the strict 2D limit we obtain the
energy of the hole-like skyrmion $E_{\mbox{\tiny
ED}+}=\frac{3}{4}\sqrt{\pi/2} r_{\mbox{\tiny C}} \hbar\omega_c$,
and the energy of the electron-like anti-skyrmion $E_{\mbox{\tiny
ED}-}= \Delta_{\mbox{\tiny ED}} -E_{\mbox{\tiny ED}+}$,
where $\Delta_{\mbox{\tiny ED}}$ is the skyrmion creation gap:
$$
\Delta_{\mbox{\tiny ED}}= \hbar\omega_c{\sqrt{\frac{\pi}{2}}
r_{\mbox{\tiny C}}}\left/\left({2+ \sqrt{\frac{\pi}{2}}
r_{\mbox{\tiny C}}}\right)\right.
$$
$$
\equiv\frac{\hbar\omega_c(e^2/\varepsilon l_B)}
{\sqrt{8/\pi}\hbar\omega_c+e^2/\varepsilon l_B}\,.   \eqno (1.2)
$$
In the EDH model the gap thereby is determined by the smallest
value among the cyclotron and Coulomb energies. At the same time,
the EDH approach gives exact results of the first order in the
expansion in terms of $r_{\mbox{\tiny C}}$. In the limit of
$r_{\mbox{\tiny C}}\to 0$ they are curiously in agreement with the
results obtained within the HF and WFP
approximations\cite{fe94,by96} when the gap is
$\frac{1}{2}\sqrt{\pi/2} r_{\mbox{\tiny C}}\hbar\omega_c$. This
coincidence of the $r_{\mbox{\tiny C}}\!\to 0\!$ results seems to
be connected only with a special symmetry of the system studied
(see the discussion in Appendix II).

The rest of the terms of the total Coulomb Hamiltonian are responsible for
other various LL mixing processes and provide
additional corrections of the second and higher degrees of
$ r_{\mbox{\tiny C}}$.
The calculation carried out in Sec. IV (at $\nu=1$) yields the exact
second-order
corrections $E^{(2)}_+=-0.008 r_{\mbox{\tiny C}}^2\hbar\omega_c$ for the
skyrmion and
$\Delta E^{(2)}=-0.382 r_{\mbox{\tiny C}}^2\hbar\omega_c$ for the
creation gap.
These corrections are independent of the magnetic field.

It is interesting that in the opposite case, when $r_{\mbox{\tiny
C}}\gg 1$, the Coulomb terms not involved in the EDH give again only
small corrections of the order of $\hbar\omega_c/r_{\mbox{\tiny
C}}$ to the fermion creation gap (see the end of Sec. IV).
Therefore, if we formally consider the $r_{\mbox{\tiny C}}\to
\infty$ limit (by keeping the magnetic field constant we can study the
$\varepsilon \to 0$ limit), then for an ideally clean sample the EDH
formula (1.2) represents also the correct result
$\Delta_{\mbox{\tiny ED}} \to \hbar\omega_c$. This has to take
place for skyrmions  which are characterized by a smooth spatial
function on the length scale of $l_B$.  The reasoning explaining
this outcome may be as follows. Indeed, at a fixed $\nu=1$ the
parameter $r_{\mbox{\tiny C}}$ has one more meaning: it is the
average inter-particle  separation in units of the effective Bohr
radius $a_B=\hbar^2 \varepsilon/m_e^*e^2$. When $r_{\mbox{\tiny
C}}\gg 1$, the average Coulomb interaction is weak as compared to
the effective Bohr atomic energy $E_B=m_e^*e^4/\hbar^2
\varepsilon^2$, and it ceases being responsible for the gap
determined by lowest-energy spatial excitations. \cite{foot0}
However, in accordance with the Kohn theorem, \cite{ko61} the
cyclotron frequency remains  always a relevant parameter of the
clean system. If $r_{\mbox{\tiny C}}\gg 1$, then $\hbar\omega_c$
is the smallest quantity in the energy scale, $\hbar\omega_c\ll
e^2/\varepsilon l_B\ll E_B$, and the fermionic gap has to approach
$\hbar\omega_c$ or zero. Since the $\nu=1$ condition seems always
to provide an insulator phase of the system studied, the result
$\Delta\to\hbar\omega_c$ is natural in the considered limit.

Of course, these simple speculations ignore a disorder which turns
out to be the main factor, and it really governs the spectrum of
the system just at $r_{\mbox{\tiny C}}\gg 1$. This case is
practically realized at comparatively low magnetic field. Then the
picture at large $r_{\mbox{\tiny C}}$ is determined by the
competition between disorder and magnetic field or between
disorder and interaction, and it turns out to be rather diverse
(e.g. see Refs. \onlinecite{hu00,chki99} and the works cited
therein).

Meanwhile, in the clean limit just the situation when
$r_{\mbox{\tiny C}}\sim 1$ becomes
experimentally relevant. Then our calculations within the EDH framework
(Sec. III) as well as beyond of it (Sec. IV)
demonstrate that the skyrmion-creation gap is
substantially reduced in comparison with the HF and
WFP calculations. In extrapolating to the region $r_{\mbox{\tiny C}}\sim 1$,
we can compare our results with the
experimental data (Sec. V), if only for the highest magnetic fields
attainable.

We note that a mean field study of the $r_{\mbox{\tiny C}}\sim
1$ ferromagnet was carried out numerically in the recent work.
\cite{mi00} The statement of the problem used there seems to
correspond to the EDH model, and the dependences found of the
skyrmion energy on $r_{\mbox{\tiny C}}$ reveal qualitatively the
same trend as in the present paper. Unfortunately, a direct
quantitative comparison with our results is impossible because in
Ref. \onlinecite{mi00} the authors report only the results for
finite 2D gas thickness or finite skyrmion spin.

To conclude the Introduction, we comment shortly on the ER method
used in the present paper. The ER technique implies a
change-over from the Fermi creation operators, which generate
one-electron eigen states of an ideal electron gas, to new exciton
operators generating one-exciton states in the 2D electron system.
When acting on the ground state, these exciton operators produce a
basic set of excitonic eigen states. An essential part of the
Coulomb interaction Hamiltonian (precisely the EDH) may be
diagonalized in this basis. We extend in this work the range of
the ER method and consider the basic set of two-exciton states. By
using a special commutation algebra of the exciton operators$\;$
\cite{by87,di96,dile99} (see also Sec. IV.A of the present paper),
ER provides a simple way to calculate matrix elements
corresponding to various types of interactions. These may be not
only the Coulomb terms (which are not involved in the EDH) but also  e.g.
electron-phonon or electron-impurity interactions. In terms of the ER,
they are renormalized respectively into
inter-exciton\cite{dile99,di99},
exciton-phonon\cite{di96,diE99,di00} or exciton-impurity
interactions. In some particular cases the ER operators were used
even in Ref. \onlinecite{dz83-84} when  studying a two-component
Fermi system with the symmetric model of interaction. Then it was
found that this model corresponds to ``inter-valley" waves in the
2D semiconductor at $\nu=1$ in a high magnetic field$\;$
\cite{by87} or to spin waves under the same
conditions.\cite{di96,di99}

$\;$

\centerline{\bf II. VARIATIONAL PRINCIPLE}

$\;$

We follow the general variational principle of
$
  E=\min\limits_{\Psi}^{}\left\{\left.
  {\left\langle\Psi|{\hat H}|\Psi\right\rangle}\right/
  {\left\langle{\vphantom{\left\langle\Psi|{\hat H}|\Psi\right\rangle}}
  \Psi|\Psi\right\rangle}\right\}\,.
$
The averaging is carried out over the sample area. If we study
an almost ferromagnetic state; i.e. the number of electrons
${\cal N}$ in the
highest
occupied LL differs from the number ${\cal N}_{\phi}$ of magnetic
flux quanta by
several units, $|{\cal N}-{\cal N}_{\phi}|\lapprox 1$,
then we can reformulate the variational principle. The desired excitation
presents
a smooth non-uniform texture determined by the rotation matrix in Eq. (1.1).
We divide the
QHF area by the great number of  $G_i$ parts, which are much
smaller than the total QHF area, but still remain much larger than the
quantum of magnetic
flux area $2\pi {}l_B^2$. The energy of excitations of
this type (including the ground state) may be found
on the basis of the minimization procedure as follows:
$$
  E=\min\limits_{U}^{}\left[\sum_i\min\limits_{\psi}^{}
  \left(\frac{\left\langle\psi|H|\psi\right\rangle_{G_i}}
  {\left\langle\psi|\psi\right\rangle_{G_i}}\right)\right], \eqno (2.1)
$$
Here, the averaging is performed over a $G_i$ area. All $G_i$ areas add up
to the total QHF area. The wave function $\psi$ should be substituted
from Eq. (1.1).

As to the outer minimization in Eq. (2.1), the only result required for its
realization is the Belavin-Polyakov theorem.\cite{be75} Let us chose a
unit vector ${\vec n}$ in the direction of the ${\hat z}'$ axis of the local
system accompanying the rotation. Evidently, $n_x=\sin{\theta}\cos{\varphi}$,
$n_y=\sin{\theta}\sin{\varphi}$, and $n_z=\cos{\theta}$, where $\varphi$ and
$\theta$ are two first Eulerian angles. The minimum of the gradient
energy in the $O(3)$ non-linear $\sigma$-model is\cite{be75}
$$
  \min\limits_{\vec n}^{}\left\{\frac{1}{2}\int d^2{\bf r}\,\left
  [\left(\partial_x{\vec n}\right)^2\,+\,
  \left(\partial_y{\vec n}\right)^2\right]\right\}\;=
  \;4\pi|q_{\mbox{\tiny T}}|\,,  \eqno (2.2)
$$
where the topological ``charge'' is
$$
  q_{\mbox{\tiny T}}=\frac{1}{4\pi}\int d^2{\bf r}\,{\vec n}
  \cdot\left(\partial_x
  {\vec n}\right)
  \times\left(\partial_y{\vec n}\right)\,.                  \eqno (2.3)
$$
This corresponds to the degree of  mapping
of the 2D plane onto a unit sphere of ${\vec n}$ directions and
therefore,
it is equal to integer number: $q_{\mbox{\tiny T}}=0,\;\pm 1,\;
\pm 2,...$.

The procedure of the inner minimization in Eq. (2.1) is equivalent to
the solution of
the Schr\"odinger equation within the area $\Delta x\Delta y=G_i$,
where the rotation is almost homogeneous,
$$
  |\partial_{\mu}U|^2\Delta x\Delta y \ll 1, \quad
  |\partial^2_{\mu}U|\Delta x\Delta y \ll 1,\quad \mu=x,y\,.  \eqno (2.4)
$$
At the same time, in zero approximation in terms of gradients, the
``regional" state is a ferromagnetic with a great number of electrons
corresponding to a local magnetic flux number
$$
  N_{\phi}({\bf r})={\Delta x\Delta y}/{2\pi {}l_B^2}\gg 1\,.
$$
For every region $G_i$ the first and second  derivatives
$\partial_{\mu}U$ and $\partial^2_{\mu\nu}U$ should be considered as
external parameters which depend only on the position of $G_i$ (e.g.
{\bf r} is the position of the center of $G_i$).

The substitution (1.1) into the Hamiltonian is reduced to a trivial 
replacement
of $\psi$ with $\chi$ in its Coulomb part, but the one-electron part becomes
of the form: \cite{io99}
$$
  {\hat H}_1=\frac{1}{2m_e^*}\int\,d^2{\bf r}\,
  {\vec \chi}^+({\bf r})
  \left(-i\nabla+{\bf A}+{\vec \Omega}^l\sigma_l\right)^2
  {\vec \chi}({\bf r}){}\qquad{}
$$
$$
  {}\qquad{} (\hbar=e/c=1,\quad l=x,y,z)\,,                                                     \eqno  (2.5)
$$
where $\sigma_{x,y,z}$ stands for Pauli matrices, and the parameters
${\vec \Omega}^{x,y,z}({\bf r})$ are proportional
to small gradients: \cite{foot2}
$$
  \Omega^z_{\mu}=\frac{1}{2}\left(1+\cos{\theta}\right)\,
  \partial_{\mu}\varphi\,,
$$
$$
  \Omega^x_{\mu}=\frac{1}{2}\left(\sin{\theta}\cos{\varphi}\,\partial_{\mu}
  \varphi-\sin{\varphi}\,\partial_{\mu}\theta\right)\,,
$$
$$
  \Omega^y_{\mu}=
  \frac{1}{2}\left(\sin{\theta}\sin{\varphi}\,
  \partial_{\mu}\varphi+\cos{\varphi}\partial_{\mu}\theta\right)\,. 
\eqno (2.6)
$$
If we were to restrict our study to the one-particle approximation and
neglect the Zeeman coupling, then the additional gauge field in Eq. (2.5)
will not give any corrections to
the one-electron energy. (Indeed, in this case
one can turn every spin in any way without any change of energy.) However, 
this
field changes effectively the ``compactness'' of the one-electron state at a
certain LL, because an additional  ``magnetic field''
$
  \nabla\times\left({\vec \Omega}^l\sigma_l\right)
$
appears.
For electrons belonging to the upper occupied LL
(at the filling $\nu=2m+1$ it has the index $m$) this additional field is
$$
  \delta{\tilde B}=\left\langle{\vec \chi}\left|\nabla\times\left(
  {\vec \Omega}^l\sigma_l\right)\right|{\vec \chi}\right\rangle_{G_i}=
   \nabla\times{\vec \Omega}^z\,, \eqno (2.7)
$$
and using Eqs. (2.6) we find also that it is equivalent to
$$
  \delta{\tilde B}=
  -\frac{1}{2}{\vec n}\cdot\left(\partial_x{\vec n}\right)
  \times\left(\partial_y{\vec n}\right)\,.         \eqno (2.8)
$$
The number of states within a LL is determined
exactly in terms of one-electron wave functions. This value is changed by
$\delta N_{\phi}=\Delta x\Delta y\delta{\tilde B}/2\pi$ for the level $m$
and the
total number of states is changed by
$\int d^2{\bf r}\delta{\tilde B}/2\pi$. Finally, due
to the principle of maximum filling, the topological invariant (2.3)
takes
on a new meaning microscopically:\cite{so93,mo95,by96,io99} $q_{\mbox{\tiny T}}$
is the number of
deficient ($q_{\mbox{\tiny T}}>0$) or excessive
($q_{\mbox{\tiny T}}<0$) electrons, i.e.,
$$
  {\cal N}={\cal N}_{\phi}-q_{\mbox{\tiny T}}\,,
  \quad\mbox{where}\quad
  {\cal N}_{\phi}=\sum_{\bf r}N_{\phi}({\bf r})\,.  \eqno (2.9)
$$

$\;$

\centerline{\bf III. MODEL OF THE ``EXCITONICALLY}
\centerline{\bf DIAGONALIZED''
COULOMB}
\centerline{\bf HAMILTONIAN AND THE FIRST ORDER} 
\centerline{\bf APPROXIMATION IN
\mbox{\large{\boldmath $ r_{\mbox{\tiny C}}$}} FOR}
\centerline{\bf THE FILLING FACTOR \mbox{\large{\boldmath $\nu=2m+1$}}}

$\;$

The additional field in the Hamiltonian (2.5)
determines a certain perturbation operator ${\hat V}_{\Omega}$, and we
can present the full Hamiltonian of the  $G_i$ region in the following form:
$$
  {\hat{\cal H}}_G={\hat H}_{01}+{\hat H}_{\mbox{\scriptsize int}}+
  {\hat V}_{\Omega}
  \,,       \eqno (3.1)
$$
where
$$
  {\hat V}_{\Omega}=\frac{1}{2m_e^*}\int\limits_{\mbox{\scriptsize over}\:
  G_i}d^2{\bf r}\,
  {\vec \chi}^+\left[
  \left(-i\nabla+{{\bf A}}+{\vec \Omega}^l\sigma_l\right)^2\right.{}\qquad{}
$$
$$
  \left.{}\qquad{}\qquad{}\qquad-\left(-i\nabla+{{\bf A}}\right)^2\right]
  {\vec \chi}\,,                                     \eqno  (3.2)
$$
and
$$
  {\hat H}_{\mbox{\scriptsize int}}=
  \lefteqn{\int}\!\int\limits_{{\mbox{\scriptsize over}\:G_i}}d^2{\bf r}
  d^2{\bf r}' {\vec \chi}^+({\bf r}){\vec \chi}^+({\bf r}'){}\qquad{}\qquad
$$
$$
 {}\qquad{}\qquad{}\cdot U({\bf r}-{\bf r}')
  {\vec \chi}({\bf r}'){\vec \chi}({\bf r})\,,     \eqno (3.3)
$$
However, by using ${\hat V}_{\Omega}$ for a perturbation technique,
we should
be accurate in avoiding a
situation where we would be solving the Sr\"odinger equation with different 
numbers of
electrons for perturbed and unperturbed parts of the Hamiltonian.
(Indeed, the number of electrons depends on
$q_{\mbox{\tiny T}}$ and therefore, on the
perturbation term.) We will solve the problem
at a fixed
$q_{\mbox{\tiny T}}$.Thus even for results associated with the
interaction
part ${\hat H}_{\mbox{\scriptsize int}}$
of the unperturbed Hamiltonian, we have to take into account that the
magnetic
field is changed effectively by the value (2.7) and (2.8) for electrons within
the LL $m$, and therefore the effective magnetic length for
this level is
$$
  {\widetilde {l_B}}={}l_B-{}l_B^3\nabla\times{\vec \Omega}^z/2\,. \eqno (3.4)
$$
At zero approximation in terms of
${\hat V}_{\Omega}$, the ground state of this $G_i$ region presents itself
as the QHF with a total spin $\frac{1}{2}{\tilde N}_{\phi}$
aligned along the ${\hat z}'$ axis of the local system,
where ${\tilde N}_{\phi}
=\Delta x\Delta y/2\pi{\widetilde {l_B}}^2$.
The Coulomb interaction does
not change the spin of the ground state.
If writing
$$
  {\hat H}_{\mbox{\scriptsize int}}={\hat H}_{\mbox{\scriptsize ED}}+
  {\hat {\cal H}}_{\mbox{\scriptsize int}}\,,  \eqno (3.5)
$$
we extract from the Coulomb Hamiltonian the  well-studied ED part
(e.g., see Refs. \onlinecite{ka84,lo93}, and the next Section of
the present paper) and will ignore so far the ${\hat {\cal
H}}_{\mbox{\scriptsize int}}$ terms. The ground state of the
Hamiltonian ${\hat H}_{01}+{\hat H}_{\mbox{\scriptsize ED}}$ is
the same state $|0\rangle$ as it is for $ r_{\mbox{\tiny C}}=0$.
The ground state energy $E_{0m}$ is determined exactly. In the
case of $\nu=1$ this energy is proportional to ${\tilde N}_{\phi
{\bf r}}e^2/\varepsilon{\widetilde {l_B}}$. Therefore the
appropriate correction (within the ED approximation) is$\;$
\cite{io00}
$$
  \delta E_{00}=\frac{3}{2}E_{00}{}l_B^2
  \nabla\times{\vec \Omega}^z\quad\mbox{for}
  \quad \nu=1\,. \eqno (3.6)
$$
(Here $E_{00}$ is considered after subtraction of the positive
background energy.)

In the case of $\nu=2m+1\ge 3$ the appropriate analysis reveals
that we have to take into account the correction (3.4) only in
${\hat H}_{\mbox{\scriptsize ED}}$ terms associated with the
single level $m$, and also in the terms responsible for the
exchange interaction between electrons of the $m$-th level and
electrons of other filled levels having the same spin state. The
result is
$$
  \delta E_{0m}=
  - \frac{3}{2}{{N}}_{\phi}(e^2/\varepsilon)
 {}l_B\nabla\times{\vec \Omega}^zI_m\,,  \eqno (3.7)
$$
where
$$
 I_{m}=\int \frac{d^2{\bf q}}{(2\pi)^2}
  V({q})e^{-q^2/2}P_{m}(q^2/2)\,. \eqno (3.8\mbox{a})
$$
Here $V({q})$ is the dimensionless 2D Fourier component of the
Coulomb potential [in the ideal 2D case we have $V({q})=2\pi/q$,
(here and everywhere below ${\bf q}$ is measured in units of
$1/{{}l_B}$)], and $P_{m}(z)$ is a polynomial of the $2m$ power.
If we set $P_0=1/2$, then the formulae (3.7) and (3.8a) determine the
energy $E_{00}$ and the correction (3.6). For $m\ge 1$, we get
$$
  P_m=\frac{1}{2}\left[L_m(z)\right]^2+\sum_{k=0}^{m}\frac{k!}{m!}z^{m-k}
  \left[L_k^{m-k}(z)\right]^2 \qquad{}\qquad{}\qquad{}\qquad{}
$$
$$
  \equiv \frac{1}{2}\left[L_m(z)\right]^2+1+
  \sum_{k=1}^{m}L_k(z)\left[L_k(z)-L_{k-1}(z)\right]\,,   \eqno
                                                       (3.8\mbox{b})
$$
where $L_k^j$ is a generalized Laguerre polynomial. [The simple
derivation for Eqs. (3.7)-(3.8a,b) may be carried out in terms of
ER by means of Eqs. (4.12)-(4.14) presented below the diagonal
part of the ${\hat H}_{\mbox{\scriptsize ED}}$ Hamiltonian.]

To find the perturbation term ${\hat V}_{\Omega}$ we substitute the expansion
${\chi}=\sum_{a p}c_{a p}{ \phi}_{a p}$ into Eq. (2.5),
where we chose the Landau-gauge
functions
as a basic set of functions $\phi_{a p}$. The subscript $p$ distinguishes the
different members of the
degenerate set of states, and the label $a$ is a binary index
$$
  a=\left(n_a,\sigma_a\right)\,,         \eqno (3.9\mbox{a})
$$
which represents both the LL index  and spin index.
Another designation will also be used when
$n$ or $\overline{n}$ is exploited as a sublevel index. In such a situation
it means that
$$
  n\equiv (n,\uparrow),\:\: \mbox{and}\:\:
  \overline{n}\equiv (n,\downarrow)\,.
                                            \eqno (3.9\mbox{b})
$$
By integrating over the
$G_i$ area in Eq. (3.2) we should substitute $x+\xi_1$ and $y+\xi_2$ for
${\bf r}$-components and $\Omega^l+
\xi_1\partial_x\Omega^l+\xi_2\partial_y\Omega^l$ for
$\Omega^l({\bf r})$, and then perform integration
over $\xi_1$ and $\xi_2$. After routine treatment we obtain
$$
  {\hat V}_{\Omega}\approx
   {\omega}_c\left({\hat {\cal U}}+{\hat {\cal U}}^++
   {\hat {\cal W}}+{\hat {\cal W}}^+\right)\,,
                                                       \eqno (3.10)
$$
where
$$
  {\hat {\cal U}}=\frac{l_B^2}{4}\left[\sum_{l}\left({\vec \Omega}^l\right)^2
  \right]{\hat {\cal N}}+
  \frac{l_B^2}{2}\nabla\!\times\!{\vec \Omega}^z
  \sum_n\left(n+\frac{1}{2}\right)\left[{\hat N}_n\right.
$$
$$
 {}\qquad{}\qquad{}{}\qquad{}\left.
 -{\hat {N}}_{\overline{n}}\right] +l_B\Omega_-^z\left({\hat K}_{\uparrow}^+
  -{\hat K}_{\downarrow}^+\right),  \eqno (3.11)
$$
and
$$
  {\hat {\cal W}}=\sqrt{N_{\phi}}\sum_n\sqrt{n+1}l_B\left[\Omega_-^-
  {\hat Q}^+_{{\overline n}\,n\!+\!1}+\Omega_-^+
  {\hat Q}^+_{n\,{\overline{n\!+\!1}}}
  \right]\,.               \eqno (3.12)
$$
The following notation is used:
$$
  \Omega^l_{\pm}=\mp\frac{i}{\sqrt{2}}
  \left(\Omega_x^l\pm i\Omega^l_y\right),
  \:\:\Omega_{\mu}^{\pm}=\left(\Omega^x_{\mu}\pm i\Omega^y_{\mu}\right);
$$
$$
 {\hat {\cal N}}= {\hat {\cal N}}_{\uparrow}+ {\hat {\cal N}}_{\downarrow}=
  \sum_n\left({\hat N}_{n}+
  {\hat N}_{\overline n}\right)\,,
$$
\mbox{where}
$$
  {\hat N}_{n}=\sum_p{\hat c}^+_{np}{\hat c}_{np}\,,\qquad
  {\hat N}_{\overline n}=
  \sum_p{\hat c}^+_{{\overline n}p}{\hat c}_{{\overline n}p}\,;
$$
$$
  {\hat K}^+_{\downarrow}=
  \sum_{n,p}\sqrt{n+1}
  {\hat c}^+_{{\overline {n\!+\!1}}p}{\hat c}_{{\overline n}p}\,,
$$
$$
  \!{\hat K}^+_{\uparrow}\!=\!
  \sum_{n,p}\sqrt{n+1}
  {\hat c}^+_{{n\!+\!1}p}{\hat c}_{np};\quad
  {\hat Q}^+_{ab}\!=\!N_{\phi}^{-1/2}
  \sum_p {b}_{p}^+{a}_{p}.\eqno (3.13)
$$
We employ here the designation ${a}_p$ (${b}_p$, ${c}_p$,...)
for the electron-annihilation operator corresponding to sublevel
$a$, ($b$, $c$,...).

The sign of approximate equality in Eq. (3.10) means that we have
omitted in the expression for ${\hat {\cal U}}$ terms of the form
$F_+\sigma_++F_-\sigma_-$
[where $\sigma_{\pm}=(\sigma_x\pm i\sigma_y)/2$, the factors $F_{\pm}$
are of the order of $(l_B\nabla)^2$]. These terms are responsible for
the deviation of all spins as a unit about the ${\hat z}'$-direction,
and they do not result in any contribution to the energy in
absence of the Zeeman coupling. \cite{foot3}

The second sum in Eq. (3.11) corresponds to the formal change of the cyclotron
energy due to the renormalization (2.7). Both operators
${\hat {\cal N}}$ and ${\hat K}^+=
{\hat K}^+_{\uparrow}+{\hat K}^+_{\downarrow}$
commute with the interaction Hamiltonian (3.3).
(This feature of ${\hat K}^+$
is a
corollary of the Kohn theorem\cite{ko61}; and in addition we have
$\left[{\hat K}^+_{\uparrow,\downarrow},{\hat H}_{01}\right]=
\omega_c {\hat K}^+_{\uparrow,\downarrow}$.) In case
${\bf |0\rangle}$ is the exact QHF ground state, then
$\left({\hat H}_{01}+{\hat H}_{\mbox{\scriptsize int}}\right){\bf |0\rangle}=
E_0{\bf |0\rangle}$; and one finds also that
$\left({\hat {\cal N}}_{\uparrow}-
{\hat {\cal N}}_{\downarrow}\right){\bf |0\rangle}=
{\tilde N}_{\phi}{\bf |0\rangle}$.

First, we
consider the correction determined by the ${\hat {\cal U}}$
terms in Eq. (3.10).
Suppose that we have the $\nu=1$ filling. In this case
${\hat { N}}_{\overline{n}}{\bf |0\rangle}=
{\hat K}^+_{\downarrow}{\bf |0\rangle}=0$. The correction
determined by ${\hat {\cal U}}$ operators in Eq. (3.10) can
be written in {\it any
order} of $ r_{\mbox{\tiny C}}$ in the general form
$$
\begin{array}{c}
  {\displaystyle {\delta E}_{{\cal U}}=
  \frac{ \omega_cl_B^2}{2}\left[\sum_{l}\left({\vec \Omega}^l\right)^2+\nabla
\times{\vec \Omega}^z\right]
  {\bf \langle 0|}{\hat{\cal N}}{\bf |0\rangle}{}\qquad{}\qquad{}{}\qquad{}}\\
   {\displaystyle +\,\omega_cl_B^2\nabla\times{\vec \Omega}^z\sum_{n=1}^{\infty}
  \,n{\bf \left\langle 0\right|}{\hat N}_{n}
   {\bf \left|0\right\rangle}}{}\qquad{}\qquad{}\\
   {\displaystyle +\,(\omega_cl_B)^2
  N_{\phi}\Omega_+^z\Omega_-^z\frac{ \left|{\bf
  \langle
  +|}{\hat K}_{\uparrow}^+{\bf |0\rangle}\right|^2}{
  E_0-E^+}\,.\;{}\;{}}\quad                                  (3.14)
  \end{array}
$$
Here ${\bf {|+ \rangle}}=
{\hat K}_{\uparrow}^+{\bf {|0 \rangle}}$ is the eigen state
 of the total unperturbed Hamiltonian (the corresponding energy is
$E^+=E_0+\omega_c$). Evidently, the first and third terms in Eq. (3.14) will
always give a
correction independent of $ r_{\mbox{\tiny C}}$. The second term also does not
result in
any
correction of the first order in  $ r_{\mbox{\tiny C}}$. [This sum gives only
corrections
of higher powers of  $ r_{\mbox{\tiny C}}$ which appear due to terms
${\hat {\cal H}}_{\mbox{\scriptsize int}}$ in the Hamiltonian (3.5).]

The  desired correction proportional to $r_{\mbox{\tiny C}}$ is thereby
determined only by the operators
${\hat {\cal W}}$ and ${\hat {\cal W}}^+$ in the perturbation (3.10).
At the same time, by operating on the ground state $|0\rangle$,
the terms
${\hat {\cal H}}_{\mbox{\scriptsize int}}$, as well as the terms of
the operator
${\hat {\cal H}}_{\mbox{\scriptsize int}}\times{\hat V}_{\Omega}$,
raise the cyclotron energy. Hence, the procedure of the perturbation theory
in  terms of ${\hat {\cal H}}_{\mbox{\scriptsize int}}$ would give only
second- or higher-order contributions to the
energy in terms of $ r_{\mbox{\tiny C}}$.

Now, let us consider $\nu=2m+1>1$. By a similar analysis we can
see that the operator (3.11) results in a contribution independent
of the Coulomb interaction or leads to other corrections which
are of the order of $ r_{\mbox{\tiny C}}^2$ and of higher orders in terms of
$ r_{\mbox{\tiny C}}$. These corrections
appear only on account of the
${\hat {\cal H}}_{\mbox{\scriptsize int}}$ terms. If we
restrict our study
to the EDH model, then ${\bf {|0 \rangle}}\equiv|0\rangle$, and each of
the operators ${\hat K}_{\uparrow}^+$ and
${\hat K}_{\downarrow}^+$ commutes by itself with
${\hat H}_{\mbox{\scriptsize ED}}$. These operators create the degenerate
state ${\hat K}_{\uparrow,\downarrow}^+|0\rangle$ with
energy $E_0+\omega_c$. Thus, if we want to solve the problem to
the first order in $ r_{\mbox{\tiny C}}$, and/or remain within the
frameworks of the
EDH model,
then we may use the ${\hat {\cal U}}$ terms only to obtain the
zeroth order contribution to the final result. (Such contributions from
all terms of
${\hat V}_{\Omega}$ cancel each other
in the zeroth order of $ r_{\mbox{\tiny C}}$.)

In this section, we consider  the EDH as a Hamiltonian responsible
for the Coulomb interaction.  As we have seen
the operator (3.12)
should really be taken into account as a perturbation. Moreover,
{\it only} operators ${\hat Q}^+_{n\,{\overline{n\!+\!1}}}$ with
$n={m\!-\!1}$ and $n=m$
have non-vanishing results of operation on $|0\rangle$ after their
commutation with
${\hat H}_{\mbox{\scriptsize ED}}$.
Therefore, only the last term of the operator ${\hat{\cal W}}$ contributes
to the energy of skyrmions.

$\;$

\centerline{\bf A.  Filling factor \mbox{\boldmath $\nu=1\quad
(m=0)$}}

$\;$

In this case, the state
$$\left|\mbox{SF}\right\rangle={\hat Q}^+_{0\,{\overline{1}}}
\left|0\right\rangle                \eqno (3.15)
$$
is an eigen state of the unperturbed Hamiltonian which corresponds
to the so-called ``spin-flip magnetoplasma'' mode$\;$
\cite{lo93,pi92} with a zero wave vector:
$$
  \left[{\hat H}_{01}+{\hat H}_{\mbox{\scriptsize ED}},
  {\hat Q}^+_{0\,{\overline{1}}}\right]
  \left|0\right\rangle=\left({\omega}_c+
  {\cal E}^{\mbox{\tiny SF}}_{01}\right)|\mbox{SF}\rangle\,.
$$
Here,  ${\cal E}^{\mbox{\tiny SF}}_{01}$ is the Coulomb part of the
energy:
$$
  {\cal E}^{\mbox{\tiny SF}}_{01}=(e^2/\varepsilon{{}l_B})
  \int \frac{d^2{\bf q}}{8\pi^2}q^2
  V({q})e^{-q^2/2}\: \eqno (3.16)
$$
If one sets formally ${\cal E}^{\mbox{\tiny SF}}_{01}=0$, then the
first- and second-order corrections of the one-electron energy, in
terms of the perturbation ${\hat V}_{\Omega}$, are exactly
canceled in the result. [One can check this fact with the  help of
Eq. (3.14) and by employing the useful identity \cite{io99} of
$\nabla\times{\vec \Omega}^i=2e_{ijk}{\vec \Omega}^j\times{\vec
\Omega}^k$.] Thus, we obtain the second-order correction
determined by the $\omega_c{\hat {\cal W}}$ perturbation:
$$
  \delta E_{{\cal W}^2}
  =\omega_c^2\frac{\Delta x\Delta y}{2\pi}\Omega_-^+\Omega_+^-
  \left|\left\langle \mbox{SF}\right|
  {\hat Q}^+_{0\,{\overline{1}}}\left|0\right\rangle\right|^2
  \left(\frac{1}{\omega_c}\right.
$$
$$
  {}\qquad{}\qquad{}\qquad{}\qquad {}\qquad{}\left.-
  \frac{1}{\omega_c+{\cal E}^{\mbox{\tiny SF}}_{01}}\right)\,.
  \eqno (3.17)
$$
The  factor $\Omega_-^+\Omega_+^-$ can be expressed in the terms of the unit
vector ${\vec n}$, since
$$
  \Omega_{\mp}^+\Omega_{\pm}^-\equiv\frac{1}{8}
  \left[\left(\partial_x{\vec n}\right)^2+\left(\partial_y{\vec n}\right)^2
  {\mp}2{\vec n}\cdot\left(\partial_x{\vec n}\times\partial_y{\vec n}\right)
  \right]\,.                             \eqno (3.18)
$$
After integration over the 2D space ($d^2{\bf r}=\Delta x\Delta
y$) we obtain with help of Eqs. (2.1)-(2.3) and (3.6)-(3.8a,b) the
skyrmion energy corresponding to the charge $q_{\mbox{\tiny T}}$:
$$
  E_{\mbox{\tiny {ED}}}(q_{\mbox{\tiny T}})=\sum_{\bf r}
  \left(\delta E_{{\cal W}^2}+\delta E_0\right)=
  \frac{{\cal E}^{\mbox{\tiny SF}}_{01}(|q_{\mbox{\tiny T}}|-
  q_{\mbox{\tiny T}})}{2\left(1+
  {\cal E}^{\mbox{\tiny SF}}_{01}/\omega_c\right)}
$$
$$
  {}\qquad{}\qquad{}\qquad{}\qquad+\frac{3e^2}
  {2\varepsilon {}l_B}I_0q_{\mbox{\tiny T}}\,.  \eqno (3.19)
$$
Therefore at $\left|q_{\mbox{\tiny T}}\right|=1$ the creation gap
is equal to
$$
  \Delta_{\mbox{\tiny {ED}}}=\frac{{\cal E}^{\mbox{\tiny SF}}_{01}}
  {\left(1+
  {\cal E}^{\mbox{\tiny SF}}_{01}/\omega_c\right)}   \eqno (3.20)
$$
(this is the factor before $\left|q_{\mbox{\tiny T}}\right|/2$).
Within the strict 2D limit, when  
$V(q)\to {2\pi}/{q}$, we get
$$
{\cal E}^{\mbox{\tiny SF}}_{01}= \frac{e^2}{\varepsilon
{}l_B}I_0=\frac{e^2}{2\varepsilon {}l_B} \sqrt{\frac{\pi}{2}}\,,
$$
and we arrive then at the result in Eq. (1.2).

$\;$

\centerline{\bf B.  Filling factor \mbox{\boldmath $\nu=2m+1\quad
(m\ge 1)$}}

$\;$

In this case, there are two basis states ${\hat Q}^+_{m\!-\!1\,
{\overline{m}}}
\left|0\right\rangle$ and ${\hat Q}^+_{m\,{\overline{m\!+\!1}}}
\left|0\right\rangle$ forming spin-flip magnetoplasma modes (with
zero wave vector) of the unperturbed
Hamiltonian. In the $2\times 2$ matrix equation of
$$
  \left[{\hat H}_{\mbox{\scriptsize ED}},{\hat Q}^+_{j\,{\overline{j\!+\!1}}}
  \right]|0\rangle=\frac{e^2}{\varepsilon {}l_B}\sum_{k=m-1}^m{\cal E}^{(0)}_{jk}
  {\hat Q}^+_{k\,{\overline{k\!+\!1}}}|0\rangle\qquad 
$$
$$
  {}\quad{}\qquad {}\qquad{}\qquad(j=m\!-\!1,m)\,,
                                         \eqno (3.21)
$$
the diagonal elements are
$$
  {\cal E}^{(0)}_{jj}=\epsilon_j=
  \int\frac{d^2{\bf q}}{(2\pi)^2}\frac{q^2V(q)}{2(j+1)}e^{-q^2/2}
  \left[L_j^1(q^2/2)\right]^2
$$
$$
  {}\quad{}\qquad {}\qquad{}\qquad(j=m\!-\!1,m)\,.                                                       \eqno (3.22)
$$
The off-diagonal matrix elements are equal to each other:
$$
  {\cal E}^{(0)}_{m\!-\!1\,{\overline{m}}}=
  {\cal E}^{(0)}_{m\,{\overline{m\!+\!1}}}=
  w_{m}=-\int\frac{d^2{\bf q}}{(2\pi)^2}\frac{q^2V(q)}
  {2\sqrt{m(m+1)}}
$$
$$
  {}\quad{}\qquad{}\times e^{-q^2/2}L_{m-1}^1(q^2/2)L_{m}^1(q^2/2)\,.
$$
Two spin-flip modes thereby have states with energies ${\omega}_c+
{\cal E}^{\mbox{\tiny SF}}_{\pm}$, where
$$
  {\cal E}^{\mbox{\tiny SF}}_{\pm}=\frac{e^2}{\varepsilon {{}l_B}}
  \left[\vphantom{\sqrt{
  \left(\epsilon_{m}-\epsilon_{m\!-\!1}\right)^2/4+ w_{m}^2}}
  \left(\epsilon_{m}+\epsilon_{m\!-\!1}\right)/2 \right.{}\qquad{}\qquad
$$
$$
   {}\qquad{}\qquad{} {}\qquad{}\left.\pm\sqrt{
  \left(\epsilon_{m}-\epsilon_{m\!-\!1}\right)^2/4+ w_{m}^2}\right]
  \,.                           \eqno (3.23)
$$
If we were to neglect the values of the commutators (3.21), then again
all corrections determined by the perturbation ${\hat V}_{\Omega}$
will add up to zero. The non-zero result for the  $G_i$ region is
determined by the second-order correction of the perturbation
theory and is caused by the ${\hat {\cal W}}$ operators
$$
 \delta E_{{\cal W}^2}
  =\frac{\Delta x\Delta y}{2\pi}\Omega_-^+\Omega_+^-
  \omega_c r_{\mbox{\tiny C}}
  {}\qquad{}\qquad {}\qquad{}\quad{}
$$
$$
  \times\frac{(m+1)\epsilon_{m}+m
  \epsilon_{m\!-\!1}+ r_{\mbox{\tiny C}}(2m+1)
  \left(\epsilon_{m}\epsilon_{m\!-\!1} -w_{m}^2\right)}
  {1+ r_{\mbox{\tiny C}}\left(\epsilon_{m}+\epsilon_{m\!-\!1}\right)+
  r_{\mbox{\tiny C}}^2 \left(\epsilon_{m}\epsilon_{m\!-\!1} -w_{m}^2\right)}\,.
  \eqno (3.24)
$$
After summation of the combination
$\delta E_{{\cal W}^2 r_{\mbox{\tiny C}}}+\delta E_{0m}$ over
all such $G_i$ regions, we find the energy of the skyrmion.
Using Eqs. (2.1)-(2.3) and (3.18) we see that in the
first order in $ r_{\mbox{\tiny C}}$
it takes the form
$$
   E_{\mbox{\tiny {ED}}}(q_{\mbox{\tiny T}})
   =\frac{e^2}{\varepsilon {}l_B}\left\{\frac{
  \left(\left|q_{\mbox{\tiny T}}\right|-q_{\mbox{\tiny T}}\right)}{2}
  \left[(m+1)\epsilon_{m}+m
  \epsilon_{m\!-\!1}\right]\right.
$$
$$
  {}\qquad{}\qquad{}{}\qquad{}\qquad{}\left.+\frac{3}{2}I_{m}
  q_{\mbox{\tiny T}}\right\}\,.  \eqno (3.25)
$$
Therefore, in this $ r_{\mbox{\tiny C}}\to 0$ limit, even for
$m=1$, the skyrmion-antiskyrmion creation gap is essentially
larger than the gap for the electron-hole pair. Indeed, with help
of Eqs. (3.8a,b) and (3.22), in the ideal 2D case we can obtain
the creation gap which turns out to be equal to
$$
  \frac{e^2}{\varepsilon{}l_B}\frac{11}{8}\sqrt{\frac{\pi}{2}}\,,
$$
whereas the appropriate value for the electron-hole pair is
$$
  \frac{e^2}{\varepsilon{}l_B}\frac{3}{4} \sqrt{\frac{\pi}{2}}\,.
$$
Just the latter determines thereby the activation charge gap in
QHF at $\nu=1$. Analogously, one can prove that for any $\nu\ge 3$
the skyrmion gap found from Eq. (3.25) is larger than the
quasiparticle gap.

Thus, skyrmions are lowest-energy fermionic excitations only in the
case of the filling factor $\nu=1$.

$\;$

\centerline{\bf IV. CORRECTIONS AT \mbox{\large{\boldmath
$\nu=1$}} TO SECOND}
\centerline{\bf ORDER IN
\mbox{\large{\boldmath $ r_{\mbox{\tiny C}}$}}}

$\;$

Generally, the second-order correction to the EDH skyrmion energy
$E_{\mbox{\tiny ED}}$ is a combination of
several parts which have different origins.

First of all, when calculating the ground-state energy to zero order in
$V_{\Omega}$ (but to second order in $ r_{\mbox{\tiny C}}$), we must
again take into
account the renormalization (2.7) and (3.4). The corresponding value, being
of the order of ${N}_{\phi} r_{\mbox{\tiny C}}^2\omega_c$, is negative (as
it has to be for any
second-order correction to the ground-state energy). Therefore, the
renormalization
correction turns out negative at positive
$\delta {\tilde B}$, namely: ${\delta E}_{0 r_{\mbox{\tiny C}}^2}\sim -\
{N}_{\phi}\nabla\times{\vec \Omega}^{z} r_{\mbox{\tiny C}}^2\omega_c$. Precisely
the same form of correction we obtain for $
{\delta E}_{{\cal U}{\cal H}^2}$
which is caused by the second term in Eq. (3.14). However,
this correction is surely positive in the case of
$\nabla\times{\vec \Omega}^{z}>0$.
In the following, we will see that both corrections cancel each other:
$$
  {\delta E}_{0 r_{\mbox{\tiny C}}^2}+{\delta E}_{{\cal U}{\cal H}^2}=0\,.
                                                               \eqno (4.1)
$$

Another correction of the required order is determined by the fourth
order of the perturbation theory in terms of the sum
${\hat {\cal H}}_{\mbox{\scriptsize {int}}}+\omega_c\left({\hat {\cal W}}+
{\hat {\cal W}}\right)$. This correction
is quadratic in ${\hat {\cal H}}_{\mbox{\scriptsize {int}}}$ and in
${\hat {\cal W}}$ and should take the form
$$
  \delta E_{{\cal W}^2{\cal H}^2}=
  N_{\phi} r_{\mbox{\tiny C}}^2\omega_c\left(\eta_1\Omega^+_-\Omega^-_++
  \eta_2\Omega^-_-\Omega^+_+\right)\,            \eqno (4.2)
$$
($|\eta_1|\sim |\eta_2|\sim 1$). For the calculation of the
corrections studied
the perturbation technique can be formulated in terms of the ER.

$\;$

\centerline{\bf A. Excitonic representation}

$\;$

We proceed from the following form of the interaction
Hamiltonian (cf. Ref. \onlinecite{dile99}):
$$
  {\hat H}_{\mbox{\scriptsize {int}}}=\frac{1}{{N}_{\phi}}
  \sum_{p,p',{\bf q}\atop a,b,c,d}
  V_{bdca}({\bf q}){}\qquad{}\qquad{}{}\qquad{}\qquad{}
$$
$$
 {}\qquad{}\qquad{}\times\exp{[iq_{x}(p'-p)]}
  b^+_{p+q_y}d^+_{p'}
  c_{p'+q_y}a_{p}\,,    \eqno (4.3)
$$
where
$$
  V_{bdca}({\bf q})=
  \frac{e^2V(q)}{2\pi\varepsilon{}l_B}h_{n_bn_a}({\bf q})
  h_{n_cn_d}^*({\bf q})\delta_{\sigma_a,\sigma_b}\delta_{\sigma_c,\sigma_d}\,.
                \eqno (4.4)
$$
The function $h_{nk}({\bf q})$ is
$$
  h_{nk}({\bf q})=\left(\frac{k!}{n!}\right)^{1/2}e^{-q^2/4}
  (q_-)^{n\!-\!k}L^{n\!-\!k}_k(q^2/2)\,,
$$
\mbox{where}
$$
  q_{\pm}=\mp\frac{i}{\sqrt{2}}(q_x\pm iq_y)\,.  \eqno (4.5)
$$
In the ER we change from electron creation (annihilation)
operators to the exciton ones:\cite{dz83-84,dz91}
$$
  {\hat {\cal Q}}_{ab{\bf q}}^{+}={N_{\phi}}^{-1/2}\sum_{p}\,
  e^{-iq_x p}
  b_{p+\frac{q_y}{2}}^{+}\,a_{p-\frac{q_y}{2}}\,,
$$
$$
  {\hat {\cal Q}}_{ab{\bf q}}=
  {\hat {\cal Q}}_{ba\,-\!{\bf q}}^+\qquad (a\not= b)\,. \eqno (4.6)
$$
This is a generalization of operators (3.13) in the case of non-zero wave
vector ${\bf q}$. A one-exciton state is defined as
$$
   |ab,{\bf q}\rangle= {\hat {\cal Q}}_{ab{\bf q}}^{+}|0\rangle\,. \eqno (4.7)
$$
We will also  use the intra-LL ``displacement'' operators
$$
  {\hat A}={N_{\phi}}^{-1/2} {\hat {\cal Q}}_{aa{\bf q}}^{+}\,,  \eqno (4.8)
$$
for which, evidently the following identity takes place:
$$
  \begin{array}{l}
  {\displaystyle {\hat A}^+_{\bf q}|0\rangle}\\ 
  \phantom{000}\\
  =\left\{
  \begin{array}{lclclcl}
  {\displaystyle \delta_{0,{\bf q}}|0\rangle}&
   {\displaystyle\mbox{if}}&
   {\displaystyle n_a<m,}&
   {\displaystyle \mbox{ or }}&
    {\displaystyle n_a=m}&
   {\displaystyle \mbox{and}}&
  {\displaystyle \sigma_a=+1/2,}\\
  {\displaystyle 0}&
  {\displaystyle \mbox{if}}&
   {\displaystyle n_a>m,}&
   {\displaystyle\mbox{or}}&
  {\displaystyle n_a=m}&
  {\displaystyle \mbox{and}}&
  {\displaystyle\sigma_a=-1/2.}\\
  \end{array}
  \right.\\
  \end{array}  \eqno (4.9)                                     
$$
The commutation rules of the operators (4.6) and (4.8) present a special Lie
algebra (cf. Refs.\onlinecite{by87,di96,dile99}):
$$
  \begin{array}{l}
  {\displaystyle \left[{\hat {\cal Q}}_{ab{\bf q_1}}^{+},
  {\hat {\cal Q}}_{ab{\bf q_2}}^{+}\right]=
  \left[{\hat {\cal Q}}_{ab{\bf q_1}},
  {\hat {\cal Q}}_{ab{\bf q_2}}\right]{}\qquad{}\qquad{}}\\
 \phantom{000}\\
 {\displaystyle {}\qquad{}\qquad{}=\left[{\hat {\cal Q}}_{bc{\bf q_1}}^{+},
  {\hat {\cal Q}}_{ab{\bf q_2}}\right]=\left[{\cal Q}_{ab{\bf q_1}}^{+},
  {\hat {\cal Q}}_{cd{\bf q_2}}^{+}\right]=0}\\ 
 \phantom{000}\\
   {\displaystyle {}\qquad{} {}\qquad{}\qquad{}{}\qquad{}\qquad(a\not=b\not=c\not=d)\,,
                                        \quad{}\; 
 {\displaystyle (4.10\mbox{a})}}\\
  \end{array}
$$
$$
  \!\!\left[{\hat {\cal Q}}_{bc{\bf q_1}}^+,{\hat {\cal Q}}_{ab{\bf q_2}}^+
  \right] {}\qquad{}\qquad{}{}\qquad{}{}\qquad{}\qquad{}{}\qquad{}
$$
$$
  =\!{N}_{\phi}^{-1/2}
   e^{-i({\bf q}_1\!\times\!{\bf q}_2)_z/2}
  {\hat {\cal Q}}_{ac\,{\bf q_1\!+\!q_2}}^+\quad
   (c\ne a,b),\eqno (4.10\mbox{b})
$$
$$
  \left[{\hat {\cal Q}}_{ab{\bf q_1}},
  {\hat {\cal Q}}_{ab{\bf q_2}}^{+}\right]=
  e^{i({\bf q}_1\!\times\!{\bf q}_2)_z/2}{\hat A}_{\bf q_1\!-
  \!q_2}{}\quad
$$
$$
  {}\qquad{}-
  e^{-i({\bf q}_1\!\times\!{\bf q}_2)_z/2}{\hat B}_{\bf q_1\!-
  \!q_2}\,,                      \eqno (4.10\mbox{c})
$$
$$
  \!e^{i({\bf q}_1\!\times\!{\bf q}_2)_z/2}\left[{\hat A}_{\bf q_1},
  {\hat {\cal Q}}_{ab{\bf q_2}}\right]\!=\!-e^{-i({\bf q}_1\!\times\!{\bf q}_2)_z/2}
   \left[{\hat B}_{\bf q_1},{\hat {\cal Q}}_{ab{\bf q_2}}\right]
$$
$$
  {}\qquad{}\qquad{}{}\qquad{}\qquad{}
  =N_{\phi}^{-1}{\hat {\cal Q}}_{ab\,{\bf q}_1\!+\!{\bf q}_2}\,. 
  \eqno (4.10\mbox{d})
$$

The interaction Hamiltonian (4.3) may be rewritten in the form
$$
  {\hat H}_{\mbox{\scriptsize {int}}}=
  \frac{1}{2}\sum_{{\bf q},a,b,c,d}
  V_{bdca}({\bf q}){\hat {\cal Q}}_{ab{\bf q}}^{+}
  {\hat {\cal Q}}_{cd\,-\!{\bf q}}^{+}
$$
$$ 
  \qquad{}{}\qquad{}\qquad{}-\sum_{{\bf q},a,b\atop (n_a\le n_b)}
  V_{baba}({\bf q}){\hat B}^+_0\,.              \eqno (4.11)
$$
Now we can extract from this expression the ED part. At least these terms do
not change the cyclotron energy. (In other words, they have to
commute with the one-electron Hamiltonian ${\hat H}_{01}$.) Therefore,
to find the EDH we should consider in Eq. (4.9) only terms with
$n_a+n_c=n_b+n_d$.
A part of these constitute an operator in which the states of the type of 
Eq. (4.7)
are the eigen states. Such a diagonal part of the EDH can be written as
$$
  {\hat H}_{\mbox{\scriptsize {ED}}}^{\mbox{\tiny {di}}}=
  \sum_{a}{\hat H}_a +\sum_{a,b\atop (a\not=b,n_a\le n_b)}
  {\hat H}_{ab}\,,                            \eqno (4.12)
$$
where
$$
  {\hat H}_{a}=\frac{1}{2}\sum_{\bf q} U_{aa}(q)
  \left({ N}_{\phi} {\hat A}^+_{\bf q}{\hat A}_{\bf q}-{\hat A}_0^+\right)\,,
                               \eqno (4.13)
$$
and
$$
  {\hat H}_{ab}=\sum_{\bf q}\,\left[U_{ab}(q){N}_{\phi}
  {\hat A}^+_{\bf q}{\hat B}_{\bf q} +{\tilde U}_{ab}(q)
  \left({\hat{\cal Q}}^+_{ab{\bf q}}{\hat{\cal Q}}_{ab{\bf q}}\right.\right.
$$
$$
  {}\qquad{}\qquad{}\qquad{}\qquad\left.\left.-
  {\hat B}_0^+\right)\right]\,.        \eqno (4.14)
$$
We have used the notations $U_{ab}=V_{abba}$ and ${\tilde U}_{ab}=V_{baba}$.
One can check that for every operator (4.6) we get
$$
  \left[{\hat H}_{a}+{\hat H}_b+{\hat H}_{ab}+\!\!\sum_{c\not=a,b}
  \left({\hat H}_{ac}\!+{\hat H}_{bc}\right)\;,\,
  {\hat {\cal Q}}_{ab{\bf q}}^{+}\right]|0\rangle
$$
$${}\qquad{}\qquad{}\qquad{}\qquad{}=
  {{\cal E}}_{ab}(q){\hat{\cal Q}}_{ab{\bf q}}^{+}|0\rangle
                                          \eqno (4.15)
$$
[In particular, if $a=(j,\uparrow)$ and $b=(j\!+\!1,\downarrow)$,
then ${{\cal E}}_{ab}(0)={\cal E}_{jj}^{(0)}$, see above Eq.
(3.22).]

However, if $m>0$ and $\delta n>0$, then the EDH also involves an 
off-diagonal
part.  When operating on the
state (4.7), the  off-diagonal terms give a finite combination
of other one-exciton states. Thus,
$$
  {\hat H}_{\mbox{\scriptsize {ED}}}=
  {\hat H}_{\mbox{\scriptsize {ED}}}^{\mbox{\tiny {di}}} +
  \sum_{ab} {\hat H}^{\mbox{\tiny {off-di}}}_{ab}              \eqno (4.16)
$$
Contrary to the definition (4.12), the summation
in Eq. (4.16)  is carried out
only over the $ab$ pairs in which the sublevel $a$ is occupied and the
sublevel $b$ is empty in the state $|0\rangle$. The members of this summation
are
$$
  {\hat H}^{\mbox{\tiny {off-di}}}_{ab}=
  \sum_{c,\!d\ne a,\!b\atop n_a+n_d=n_b+n_c}\sum_{\bf q}\left[
  {V}_{adcb}(q){\hat{\cal Q}}_{cd{\bf q}}^{+} {\hat{\cal Q}}_{ab{\bf q}}\right.
$$
$$ {}\qquad{}\qquad{}{}\qquad{}+\left.
  {V}_{adbc}(q){\hat{\cal Q}}_{ca{\bf q}}^{+}
  {\hat{\cal Q}}_{db{\bf q}}\right]\,.
                              \eqno (4.17)
$$
One can check that
$$
  \!\!\left[{\hat H}^{\mbox{\tiny {off-di}}}_{ab},
  {\hat {\cal Q}}_{ab{\bf q}}^{+}\right]|0\rangle=\!\!
  \sum_{a',\!b'\ne a,\!b}
  {\cal E}_{a'b'}^{(ab)}(q){\hat{\cal Q}}_{a'b'{\bf q}}^{+}|0\rangle\,,
                                           \eqno (4.18)
$$
and the pairs of the states $a'b'$... in Eq. (4.18) provide the same
$\delta n$ and $\delta S_z$ as those in the case of
the pair $ab$:
$$
 \delta n=n_b-n_a=n_{b'}-n_{a'}=...,
$$
$$
  \delta S_z=\sigma_b-\sigma_a=\sigma_{b'}-\sigma_{a'}=...\,.     \eqno (4.19)
$$

The finite set of equations (4.15) and (4.18) thereby determines
the eigen energies and the eigen states of the EDH which
correspond to given $\delta n$, ${\delta S}_z$ and ${\bf q}$.
[Specifically, in this way the spin-flip modes at ${\bf q}=0$
(3.23) have been found.]

All other terms of ${\hat {\cal H}}_{\mbox{\scriptsize {int}}}=
{\hat H}_{\mbox{\scriptsize {int}}}-{\hat H}_{\mbox{\scriptsize {ED}}}$, with
which an operator ${\hat{\cal Q}}_{ab{\bf q}}^+$ does not commute, have the
following form:
$$
  {\hat H}_{ab}'=\sum_{c,\!d\ne a,\!b \atop g\ne a}\sum_{\bf q}
  {V}_{gacd}(q){\hat {\cal Q}}_{dg{\bf q}}^{+} {\hat {\cal Q}}_{ac{\bf q}}
  {}\qquad{}\qquad{}
$$
$$
  \qquad{}{}\qquad{}+
  \sum_{c,d\ne a,b \atop g\ne b}\sum_{\bf q}
  {V}_{cdbg}(q){\hat {\cal Q}}_{gc{\bf q}}^{+}{\hat {\cal Q}}_{db{\bf q}}\,.
                                   \eqno (4.20)
$$
If operating on the state (4.7), these terms lead to ``superfluous''
two-exciton
states. Meanwhile, some operators (4.20) do not change the cyclotron energy
and
even within the approximation of the first order in $ r_{\mbox{\tiny C}}$,
they must be considered
for the correct
calculation of exciton energy. Specifically, for the spin-flip mode
($a=0,\:b={\overline 1},$ if $\nu=1$) the terms
$$
  {\hat H}_{0{\overline 1}}'=\sum_{\bf q}V_{1010}(q)
  {\hat {\cal Q}}_{01{\bf q}}^{+}
  {\hat {\cal Q}}_{\overline{0}\,\overline{1}{\bf q}}
  \quad+\quad  \mbox{H. c.}   \eqno (4.21)
$$
also have to be taken into account as well as those of
${\hat H}_{\mbox{\scriptsize {ED}}}$.

We will calculate the second-order corrections to the energy in the
case of the
filling factor $\nu=1$. Therefore,
within the framework of our problem, we are interested in results of the
operation
of ${\hat {\cal H}}_{\mbox{\scriptsize {int}}}$ on the state $|0\rangle$ or
${\hat {\cal Q}}^+_{0\overline{1}\,{\bf 0}}|0\rangle$, and
it should be chosen in the form
$$
  {\hat {\cal H}}_{\mbox{\scriptsize {int}}}=
  \frac{1}{2}\sum_{{\bf q},n_b>0,n_d>0}
  V_{bd00}({\bf q}){\hat {\cal Q}}_{0b{\bf q}}^{+}
  {\hat {\cal Q}}_{0d\,-\!{\bf q}}^{+}{}\qquad{}\qquad{}
$$
$$
  {}\qquad{}+\sum_{{\bf q},n_b,n_d}
  V_{bd\overline{1}0}({\bf q}){\hat {\cal Q}}_{0b{\bf q}}^{+}
  {\hat {\cal Q}}_{\overline{1}d\,-\!{\bf q}}^{+}\quad +\quad \mbox{H. c.}
                                          \eqno (4.22)
$$
The terms (4.21) enter into the second sum of this expression.

$\;$

\centerline{\bf B. Two-exciton states at \mbox{\boldmath $\nu=1$}}

$\;$

The operation of the Hamiltonian (4.22) on the EDH ground state $|0\rangle$,
at $\nu=1$, leads to two-exciton states of the type of
$$
  |\alpha\rangle=
  {\hat {P}}^+_{\alpha}|0\rangle\,,   \eqno (4.23)
$$
where we will denote as ${\hat {P}}^+_{\alpha}$ the two-exciton creation
operator
$$
   {\hat {P}}^+_{\alpha}= \frac{1}{2}{\hat {\cal Q}}^+_{0a_1{\bf q}_{\alpha}}
   {\hat {\cal Q}}^+_{0a_2-\!{\bf q}_{\alpha}}\,   \eqno (4.24)
$$
and designate as $\alpha$ the composite index $\alpha=
(a_1,a_2,{\bf q}_{\alpha})$ [correspondingly
$\beta=(b_1,b_2,{\bf q}_{\beta})$,...],
which obeys the evident condition
$$
  |\alpha\rangle=|a_1,a_2,{\bf q}_{\alpha}\rangle\equiv
  |a_2,a_1,-{\bf q}_{\alpha}\rangle\,.             \eqno (4.25)
$$
Any state (4.23) is a ``quasi'' eigenstate of the unperturbed
Hamiltonian ${\hat H}_{01}+{\hat H}_{\mbox{\scriptsize {ED}}}$, since
$$
  \left[{\hat H}_{01}+{\hat H}_{\mbox{\scriptsize {ED}}},
  {\hat P}^+_{\alpha}\right]|0\rangle=
  \left[\omega_c\left(n_{a_1}+n_{a_2}\right)
  +{\cal E}_{0a_1}(q_{\alpha})\right.{}\qquad{}\qquad{}
$$
$$
  \qquad{}{}\qquad{}\qquad{}\left.+{\cal E}_{0a_2}(q_{\alpha})
  \right]
  {\hat P}^+_{\alpha}|0\rangle +
  \frac{e^2}{\varepsilon{}l_B}|{\tilde \varepsilon}\rangle \eqno (4.26)
$$
[${\cal E}_{0a_1}$ and ${\cal E}_{0a_2}$ are Coulomb energies
determined by Eqs. (4.15)], where the state
$|{\tilde \varepsilon}\rangle$ has
a norm of the order of $1/N_{\phi}$.
However, in comparison with the set of orthogonal one-exciton states (4.7),
the states (4.23) are ``slightly'' nonorthogonal to each other.
We can find that
$$
  \langle\alpha|\beta\rangle=\frac{1}{4}\left\{\delta_{a_1,b_1}
  \delta_{a_2,b_2}
  \left[\delta_{{\bf q}_{\alpha},{\bf q}_{\beta}}-\frac{1}
  {N_{\phi}}e^{i\left(
  {\bf q}_{\alpha}\!\times\!{\bf q}_{\beta}\right)_z}\right]\right.
  {}\qquad{}\qquad{}\qquad{}
$$
$$
  \left.+\,\delta_{a_2,b_1}\delta_{a_1,b_2}
  \left[\delta_{{\bf q}_{\alpha},-\!{\bf q}_{\beta}}-\frac{1}
  {N_{\phi}}e^{i\left(
  {\bf q}_{\beta}\!\times\!{\bf q}_{\alpha}\right)_z}\right]\right\}\,.
                                                    \eqno (4.27)
$$

Meanwhile, this nonorthogonality has to be taken into account if
we are to consider a combination
$$
    \sum_{\beta}\varphi_{\beta}|\beta\rangle        \eqno (4.28)
$$
(this is a summation over all of the members
of the composite index). In this case, the function
$\varphi_{\beta}=\varphi(b_1,b_2,{\bf q}_{\beta})$
turns out to be non-single-valued
one. Indeed, let us project this state onto a certain state (4.23).
We obtain
$$
  \sum_{\beta}\varphi_{\beta}\langle\alpha|\beta\rangle =
  \frac{1}{2}\left(\varphi_{\alpha}-\overline{\varphi}_{\alpha}\right)\,,
                                       \eqno (4.29)
$$
where
$$
  \overline{\varphi}_{\alpha}=\frac{1}{N_{\phi}}
  \sum_{\beta}{\cal F}_{\alpha\beta}
  \varphi_{\beta}                        \eqno (4.30)
$$
is a Fourier transform determined by the kernel
$$
  {\cal F}_{\alpha\beta}=\delta_{a_1,b_1}\delta_{a_2,b_2}e^{i\left(
  {\bf q}_{\alpha}\!\times\!{\bf q}_{\beta}\right)_z}\,.     \eqno (4.31)
$$
If $\varphi_{\alpha}=\overline{\varphi}_{\alpha}$,
then any
projection (4.29) is equal to zero. Only the ``antisymmetrized'' part
$$
  \varphi_{\beta}^{(a)}=\frac{1}{2}
  \left(\varphi_{\beta}-\overline{\varphi}_{\beta}\right)  \eqno (4.32)
$$
contributes thereby to the combination (4.28). The origin of this feature
of states (4.28) is
related to the permutation antisymmetry of the electron wave function in
the system studied (cf. for example Ref. \onlinecite{by83}). Note also that
$\overline{\overline{\varphi}}_{\alpha}=\varphi_{\alpha}$ and
$\sum_{\alpha}w(a_1,a_2){\psi_{\alpha}^*}\varphi_{\alpha}=
\sum_{\alpha}w(a_1,a_2)\overline{{\psi_{\alpha}^*}}
\overline{\varphi_{\alpha}}$, where $w$ is such that $w(a_1,a_2)=
w(a_2,a_1)$. As a result,
we get the useful equivalence
$$
  \sum_{\alpha}w(a_1,a_2){\psi_{\alpha}^*}^{(a)}\varphi_{\alpha}^{(a)}=
  \sum_{\alpha}w(a_1,a_2){\psi_{\alpha}^*}\varphi_{\alpha}^{(a)}{}\quad{}
$$
$$
  {}\qquad{}\qquad{}{}\qquad{}\qquad{}=
  \sum_{\alpha}w(a_1,a_2){\psi_{\alpha}^*}^{(a)}\varphi_{\alpha}\,.
                                                          \eqno (4.33)
$$

In terms of these definitions the expectation (4.27) may be rewritten as
$$
  \langle \alpha|\beta\rangle=\frac{1}{2}\left(\delta_{\alpha,\beta}-
  \frac{1}{N_{\phi}}\sum_{\gamma}{\cal F}_{\alpha\gamma}\delta_{\gamma,\beta}
  \right)\equiv\delta_{\alpha,\beta}^{(a)}\,,     \eqno (4.34)
$$
where
$$
  \delta_{\alpha,\beta}=\frac{1}{2}\left(\delta_{a_1,b_1}
  \delta_{a_2,b_2}
  \delta_{{\bf q}_{\alpha},{\bf q}_{\beta}}+
  \delta_{a_2,b_1}\delta_{a_1,b_2}
  \delta_{{\bf q}_{\alpha},-\!{\bf q}_{\beta}}\right)\,.       \eqno (4.35)
$$

$\;$

\centerline{\bf C. Perturbation-theory results}

$\;$

When ${\hat {\cal H}}_{\mbox{\scriptsize {int}}}+\omega_c\left({\hat {\cal W}}+
{\hat {\cal W}}^+\right)$
is a perturbation and
${\hat H}_{01}+{\hat H}_{\mbox{\scriptsize {ED}}}$ is an unperturbed
Hamiltonian, then it is sufficient to employ as a basic set the
two-exciton states (4.24)
and the spin-flip state (3.15). Thus the correction to the EDH ground
state $|0\rangle$ may be presented in the form
$$
  \delta{\bf {|0\rangle}}=C_0|0\rangle+
  \sum_{\alpha\not=0}C_{\alpha}|\alpha\rangle+
  D|\mbox{SF}\rangle\,,                              \eqno (4.36)
$$
where the factors $C_{\alpha}$ and $D$, should be found in a specified order in
terms of  ${\hat {\cal H}}_{\mbox{\scriptsize {int}}}$ and ${\hat {\cal W}}$
(actually in terms of $ r_{\mbox{\tiny C}}$ and $\Omega^l_j$). In our case,
where we are
concerned only with the antisymmetrized functions
$C_{\alpha}\equiv C_{\alpha}^{(a)},{}\;$ the above equations
(4.29), (4.33) and 

\clearpage

\onecolumn
\noindent
(4.34) show that two-exciton states may be considered as an
orthogonal and normalized basis, for which the perturbation-theory technique
may be used in its traditional form.

First, we find $C_{\alpha}$ to the first order in
${\hat {\cal H}}_{\mbox{\scriptsize {int}}}$,
$$
  C_{0{\cal H}}=0,\quad
  C_{\alpha {\cal H}}=-{\left\langle\alpha\left
  |{\hat {\cal H}}_{\mbox{\scriptsize {int}}}\right|0\right\rangle}/
{\Delta^c_{\alpha}}\quad(\alpha\not=0).   \eqno (4.37)
$$
where ${\Delta^c_{\alpha}}=\omega_c\left(n_{a_1}+n_{a_2}\right)$ stands for the
cyclotron part of energy of the two-exciton state $\left|\alpha\right\rangle$
[c.f. Eq. (4.27)]. When substituting 
${\bf |0\rangle}=|0\rangle+\sum_{\alpha}C_{\alpha {\cal H}}
|\alpha\rangle$ 
into the second term of Eq. (3.14), we obtain
${\delta E}_{{\cal U}{\cal H}^2}=\nabla\times{\vec \Omega}^{z}\omega_cl_B^2
\sum_{\alpha}\left(n_{a_1}+n_{a_2}\right)\left|C_{\alpha {\cal H}}\right|^2$.
Then by calculating
also the second order correction ${\delta E}_{0 r_{\mbox{\tiny C}}^2}=
\nabla\times{\vec \Omega}^{z}l_B^2\sum_{\alpha}C_{\alpha {\cal H}}
\left\langle 0\left
  |{\hat {\cal H}}_{\mbox{\scriptsize {int}}}\right|\alpha\right\rangle$, we
come indeed to the  result of zero in the combination (4.1).

The desired correction (4.2) is determined by means of a conventional
procedure. In which, factors $C_{\alpha}$ have to be found sequentially
up to the second order in ${\hat {\cal W}}$ and to the first order in
${\hat {\cal H}}_{\mbox{\scriptsize {int}}}$. Whereas $D$, is determined
to the second order in ${\hat {\cal H}}_{\mbox{\scriptsize {int}}}$ and
the first order in  ${\hat {\cal W}}$. The result is written in the form
$$
   \delta E_{{\cal W}^2{\cal H}^2}={\cal A}_1+2{\cal A}_2+{\cal A}_3\,,
                                              \eqno (4.38)
$$
where
$${\cal A}_1=-\omega_c^2\sum_{\alpha,\beta,\gamma}
 \frac{\left\langle 0\left
  |{\hat {\cal H}}_{\mbox{\scriptsize {int}}}\right|\gamma\right\rangle
  \left\langle 0\left|\left[{\hat P}_{\gamma},{\hat {\cal W}}^+
  \right]\right|\beta\right\rangle
  \left\langle\beta\left|\left[{\hat {\cal W}},{\hat P}_{\alpha}^+
  \right]\right|0\right\rangle
  \left\langle\alpha\left
  |{\hat {\cal H}}_{\mbox{\scriptsize {int}}}\right|0\right\rangle}
  {\Delta^c_{\gamma}\Delta^c_{\beta}\Delta^c_{\alpha}}\,,    \eqno (4.39\mbox{a})
$$
$$
  {\cal A}_2=-\omega_c^2\sum_{\beta,\gamma}
  \frac{\left\langle 0\left
  |{\hat {\cal H}}_{\mbox{\scriptsize {int}}}\right|\gamma\right\rangle
  \left\langle 0\left|\left[{\hat P}_{\gamma},{\hat {\cal W}}^+
  \right]\right|\beta\right\rangle
  \left\langle\beta\left|\left[{\hat {\cal H}}_{\mbox{\scriptsize {int}}},
  {\hat Q}^+_{0\overline{1}}\right]\right|
  0\right\rangle
  \left\langle\mbox{SF}\left|{\hat {\cal W}}\right|0\right\rangle}
  {\Delta^c_{\gamma}\Delta^c_{\beta}\Delta^c_{\mbox{\scriptsize SF}}}\,,
  \quad\mbox{and}          \eqno (4.39\mbox{b})
$$
$$
  {\cal A}_3=-\omega_c^2\sum_{\beta}\frac{
  \left\langle 0\left|{\hat {\cal W}^+}\right|\mbox{SF}\right\rangle
  \left\langle 0\left|\left[
  {\hat Q}_{0\overline{1}},{\hat {\cal H}}_{\mbox{\scriptsize {int}}}\right]
  \right|
  \beta\right\rangle
  \left\langle\beta\left|\left[{\hat {\cal H}}_{\mbox{\scriptsize {int}}},
  {\hat Q}^+_{0\overline{1}}\right]\right|
  0\right\rangle
  \left\langle\mbox{SF}\left|{\hat {\cal W}}\right|0\right\rangle}
  {\Delta^c_{\beta}\left({\Delta^c_{\mbox{\scriptsize SF}}}\right)^2}\,,
                                                       \eqno (4.39\mbox{c})
$$
(we set here $\Delta^c_{\mbox{\scriptsize SF}}=\omega_c$ which is the
cyclotron part of energy in the state $|\mbox{SF}\rangle$). The matrix elements
entering into these expressions are
calculated in Appendix I.

Consider for example, the term ${\cal A}_1$.
With help of Eqs. (A1.1), (A1.2)
(A1.5), and (A1.7) in the Appendix I and using Eq. (4.33), we find that Eq.
(4.39a) is changed into
$$
  {\cal A}_1=\!- r_{\mbox{\tiny C}}^2\omega_c N_{\phi}\left[\Omega^+_-\Omega^-_+
  \sum_{\alpha}\frac{g_{\alpha}^*g_{\alpha}^{(a)}
  \left(n_{a_1}+n_{a_2}+2\right)}{\left(n_{a_1}\!+n_{a_2}\right)^2
  \left(n_{a_1}\!+n_{a_2}\!+1\right)}
  +\Omega^+_+\Omega^-_-
  \sum_{\alpha}\frac{g_{\alpha}^*g_{\alpha}^{(a)}
  \left(n_{a_1}+n_{a_2}\right)}{\left(n_{a_1}\!+n_{a_2}\right)^2
  \left(n_{a_1}\!+n_{a_2}\!-1\right)}\right]\,.        \eqno (4.40)
$$

After substituting Eq. (A1.2) for $g_{\alpha}$, the suitable sequence of
mathematical treatments is as follows: we perform the
summation over all of $n_{a_1}\ge 1$ and  $n_{a_2}\ge 1$
keeping the sum $n_a=n_{a_1}+n_{a_2}\ge 2$ fixed;
then we  make
the integration
over ${\bf q}_{\alpha}$ [the antisymmetrized function
$g_{\alpha}^{(a)}$  already contains an integration according
to Eqs. (4.30) and (4.31); therefore the second term in the expression
$g_{\alpha}^*g_{\alpha}^{(a)}=
(|g_{\alpha}|^2-g_{\alpha}^*\overline{g}_{\alpha})/2$ leads to twofold
integration over  ${\bf q}_{\beta}$ and ${\bf q}_{\alpha}$
which, however, can be reduced analytically to a simple
onefold integral];
and finally the numerical summation over $n_a$ is performed.
In the ideal 2D
case
[i.e. if $V(q)=2\pi/q$] the result is
$$
  {\cal A}_1=- r_{\mbox{\tiny C}}^2\omega_cN_{\phi}
  \left(0.0056\Omega^+_-\Omega^-_++
  0.0077\Omega^+_+\Omega^-_-\right)\,.      \eqno (4.41)
$$

In a like manner, the calculation of ${\cal A}_2$ and ${\cal A}_3$ can be
carried out. In so doing, for the ideal 2D system limit, we obtain
${\cal A}_2=-0.0248 r_{\mbox{\tiny C}}^2\omega_cN_{\phi}\Omega^+_-\Omega^-_+$
and
${\cal A}_3=-0.318 r_{\mbox{\tiny C}}^2\omega_cN_{\phi}\Omega^+_-\Omega^-_+$.
[Notice that the
operators (4.21), which do not change the cyclotron energy, contribute only
to the term  ${\cal A}_3$.]

With help of Eq. (3.18) and Eqs. (2.1)-(2.3)
the summation over all of the $G_i$ regions yields the
 second-order correction to the
EDH result (3.19):
$$
  {E}^{(2)}_{\mbox{\it q}_{\mbox{\tiny T}}}=\sum_{G_i}
  \delta E_{{\cal W}^2{\cal H}^2}=
  r_{\mbox{\tiny C}}^2\omega_c\left[\frac{q_{\mbox{\tiny T}}}{2}
  \left(\eta_2-\eta_1\right)
  +\frac{|q_{\mbox{\tiny T}}|}{2}\left(\eta_1+\eta_2\right)\right]\,,
                                                                  \eqno (4.42)
$$
where $\eta_1=-0.374$ and $\eta_2=-0.0077$. The correction to the EDH skyrmion-antiskyrmion
creation gap is
$$
    \Delta E^{(2)}=r_{\mbox{\tiny C}}^2\omega_c \left(\eta_1+\eta_2\right)\,.  
\eqno (4.43)
$$
With Eqs. (3.19) and (4.42) we find
the total $ r_{\mbox{\tiny C}}^2$ correction to the HF{$\,$}
\cite{fe94,by96,fe97,io99,io00}
result:
$$
  E^{(2)}_{\mbox{\scriptsize {tot}}\,{\mbox{\it q}_{\mbox{\tiny T}}}}=
  -\frac{\left({\cal E}^{\mbox{\tiny SF}}_{01}\right)^2}{2\omega_c}
  \left(|q_{\mbox{\tiny T}}|-q_{\mbox{\tiny T}}\right)+
  {E}^{(2)}_{\mbox{\it q}_{\mbox{\tiny T}}}\,.
$$

\twocolumn
The corrections for quasiparticles are correspondingly:
$E^{(2)}_{\mbox{\scriptsize {tot}}\,-}=
-0.767r_{\mbox{\tiny C}}^2\omega_c$ for the electron-like antiskyrmion
and
$E^{(2)}_{\mbox{\scriptsize {tot}}\,+}=
-0.0077r_{\mbox{\tiny C}}^2\omega_c$ for the holelike skyrmion.\cite{foot4}

To conclude this section we prove that all perturbative (in terms
of ${\cal H}_{\mbox{\scriptsize int}}$) corrections to the EDH gap
(1.2) vanish in the $r_{\mbox{\tiny C}}\to \infty$ limit. Note
that the above calculations of ${\delta E}_{0 r_{\mbox{\tiny
C}}^2}$, ${\delta E}_{{\cal U}{\cal H}^2}$ and $E_{{\cal W}^2{\cal
H}^2}$ present the corrections of the second order in ${\cal
H}_{\mbox{\scriptsize int}}$ where the energies of one- and
two-exciton basis states are considered within the zero approximation
in $r_{\mbox{\tiny C}}$. At the same time, the technique used
provides a formal possibility to develop perturbatively an
expansion in ${\cal H}_{\mbox{\scriptsize int}}$ for arbitrary,
$r_{\mbox{\tiny C}}$. To do this we should replace
$\Delta_{\mbox{\scriptsize {SF}}}$ and
$\Delta^c_{\alpha,\beta,\gamma}$ in Eqs. (4.37) and Eqs. (4.39a-c)
with their values exactly calculated within the EDH model. That is
we should add the corresponding Coulomb shifts:
$\Delta_{\mbox{\scriptsize {SF}}} \to \omega_c+{\cal
E}^{\mbox{\tiny SF}}_{01}$ and $\Delta^c_{\alpha}\to
\omega_c\left(n_{a_1}\!+\!n_{a_2}\right) +{\cal
E}_{0a_1}(q_{\alpha})+{\cal E}_{0a_2}(q_{\alpha})$
[$\Delta^c_{\beta,\!\gamma}=\Delta^c_{\alpha}( a\!\to\! b,c)$, see
Eqs. (3.16) and (4.15)]. Naturally, at $r_{\mbox{\tiny C}}\lesssim
1$ this procedure becomes senseless if we were to restrict our
consideration  to the second order in ${\cal H}_{\mbox{\scriptsize
int}}$ only.  Indeed, both of the operators $ {\hat
H}_{\mbox{\scriptsize {ED}}}$ and
 ${\cal H}_{\mbox{\scriptsize int}}$ are proportional to
 $e^2/\varepsilon l_B$ and the
accounted Coulomb shifts in the denominators of Eqs. (4.37) and
(4.39a-c) would only yield  the $r_{\mbox{\tiny C}}^3$ and higher
order corrections which are beyond this approximation.

However, let $r_{\mbox{\tiny C}}\gapprox 1$, and estimate at once
{\it all terms} of the perturbative expansion in ${\cal
H}_{\mbox{\scriptsize int}}$. In this case
$\left|C_{\alpha}\right|\sim 1$ to any order in ${\cal
H}_{\mbox{\scriptsize int}}$, and now there is no cancellation
(4.1), because ${\delta E}_{0 r_{\mbox{\tiny C}}^{\kappa}}
\not=-{\delta E}_{{\cal U}{\cal H}^{\kappa}}$ for any
${\kappa}>1$. (The more specific estimations are ${\delta E}_{0
r_{\mbox{\tiny C}}^{\kappa}}\sim \nabla\!\times\!{\vec
\Omega}^zl_B^2\omega_c r_{\mbox{\tiny C}}$ and ${\delta E}_{{\cal
U}{\cal H}^{\kappa}}\sim \nabla\!\times\!{\vec
\Omega}^zl_B^2\omega_c$.) Nevertheless after integration over
2D space both of these corrections contribute only to the term
proportional to the charge $q_{\mbox{\tiny T}}$ (2.3), and
therefore they do not contribute to the creation gap of
skyrmion-antyskyrmion pares. As to corrections ${\delta E}_{{\cal
W}^2{\cal H}^{\tiny \kappa}}$, we find easily that  they are all
of the order of $l_B^2\omega_c\nabla^2/r_{\mbox{\tiny C}}$ and
give a correction  to the  gap of the order of
$$
  \sum_{\kappa \geq 2}\Delta E^{(\kappa)} \sim \omega_c/r_{\mbox{\tiny C}}\,.
$$
When $r_{\mbox{\tiny C}}\gg 1$, the EDH value
$\Delta_{\mbox{\scriptsize ED}}\approx \omega_c$
presents thereby the main  part of the creation gap.

$\;$

\centerline{\bf V. DISCUSSION}

$\;$

Thus, our calculations consist of two main stages. In the first stage
we have considered only the ED part of the Hamiltonian, where
the LL mixing is partly taken into account. The corresponding
creation gap for charged quasiparticles $\Delta_{\mbox{\tiny ED}}$
(i.e. for $|q_{\mbox{\tiny T}}|=1$
skyrmion-antiskyrmion pairs) is shown in Fig. 1.  Here,
$r_{\mbox{\tiny C}}$ is an arbitrary parameter
(formally it does not need to be
small in this approach). We see that even the EDH model reflects a
significant reduction of the gap with a
growing $r_{\mbox{\tiny C}}$.
Besides the obtained $\Delta_{\mbox{\tiny ED}}$ yields the correct
limiting values for $r_{\mbox{\tiny C}}\to 0$ and for
$r_{\mbox{\tiny C}}\to \infty$.

In the second stage we have treated the remaining part of the
Hamiltonian perturbatively in $r_{\mbox{\tiny C}}$, calculating
the correction to second order. Needless to say it would be
incorrect to apply this $r_{\mbox{\tiny C}}^2$ correction to the
case when $r_{\mbox{\tiny C}}^2\gapprox 1$. Nevertheless, the
perturbation theory result indicates at least the tendency of the
creation-gap variation with  $r_{\mbox{\tiny C}}$. Fig. 1 displays
also the corrected value
$$
  \Delta_{\mbox{\tiny ED}}^{\mbox{\tiny corr}}=\Delta_{\mbox{\tiny ED}}+
  \Delta E^{(2)}\,,                          \eqno     (5.1)
$$
where
$$
  \Delta E^{(2)}=\left(\eta_1+\eta_2\right)r_{\mbox{\tiny C}}^2\omega_c\,.
                                               \eqno (5.2)
$$
We  choose  conventionally the region $r_{\mbox{\tiny C}}<0.2$ as a region
of the small  $r_{\mbox{\tiny C}}$ values where the result (5.1) is
correct. The curve for $\Delta_{\mbox{\tiny ED}}^{\mbox{\tiny corr}}$
trends
thereby to a more severe decrease in the gap. In addition, we extend
this curve further (by the dashed line)
approximately to
$r_{\mbox{\tiny C}}$ which corresponds to the
experimental conditions \cite{ma96} (
the creation gap measurements made  in Ref.
\onlinecite{ma96} seem
to be the most high magnetic field results available to date).

\vspace{-0.8cm}

\begin{figure}[t]
\vspace{0cm}
\centerline{\psfig{figure=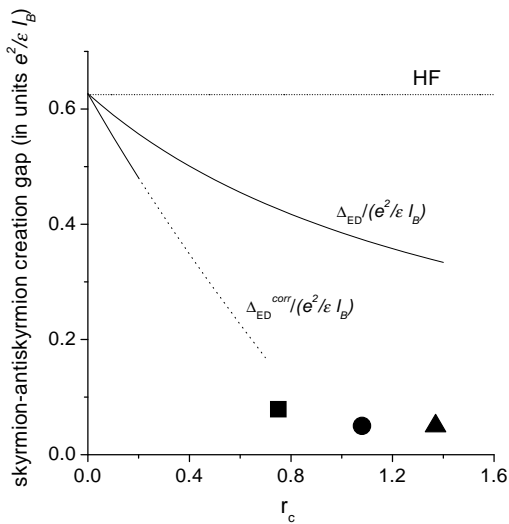,width=90mm,angle=0}}
\vspace{-4.2cm}
\caption{The normalized creation gap of the ED Hamiltonian
 model and the
corrected gap are shown above. The dotted line represents the
HF approximation result. The experimental data (closed symbols)
are from Refs. 3 (the square), 4 (the circle), and 5 (the
triangle).}
\label{Firstfig}
\end{figure}

\vspace{0.5cm}
Such results for $\Delta_{\mbox{\tiny ED}}$ and for
$\Delta_{\mbox{\tiny ED}}^{\mbox{\tiny corr}}$ demonstrate a
significant role of the LL mixing. The curves in Fig. 1 become
closer to experimental data (c.f. the HF value shown by the dotted
line). We demonstrate here  only the data found at fairly strong
magnetic fields, namely: $B=11.6,\;5.5\;\;\mbox{and} \;\;3.5\,$T
correspondingly in Refs. \onlinecite{ma96,sh00,me99}. Meanwhile,
it should be noted, that the creation gap experimentally observed
depends sharply on the effective $g$ factor in the vicinity of
$g=0$ [in accordance with the low $(g\ln g)^{1/3}$, see Ref.
\onlinecite{so93}]. The result presented in the work of Ref.
\onlinecite{ma96} and marked here by the square is really obtained
at zero Zeeman energy (to an experimental accuracy). The
measurements \cite{me99,sh00} are carried out at small, but still
not vanishing $g$ factors. Therefore, we have extrapolated these
data to the point $g=0$. The results of this conventional
extrapolation are presented in Fig. 1 by the closed triangle and
the circle .

At the same time, the dashed line for
value $ \Delta_{\mbox{\tiny ED}}^{\mbox{\tiny corr}}$
evidently presents an underestimated result. In the higher order corrections
in terms of $r_{\mbox{\tiny C}}$ the curve corresponding to the true energy gap
should pass somewhere between the curves of Fig. 1 (the smaller
the parameter
$r_{\mbox{\tiny C}}$ is, the closer this true gap should be to the
calculated value
$ \Delta_{\mbox{\tiny ED}}^{\mbox{\tiny corr}}$). In the ideal 2D case
the results $ \Delta_{\mbox{\tiny ED}}$ and
$ \Delta_{\mbox{\tiny ED}}^{\mbox{\tiny corr}}$ calculated for
$r_{\mbox{\tiny C}}\sim 1$
may be considered as upper and lower limits for a real creation gap.

Finally we should emphasize that at least two important effects
have been ignored with the calculation of the Fig. 1 curves:
finite thickness (FT) correction, and disorder broadening of
Landau levels (LL's). Both of these reduce the energy gap. The
usual way to take into account the FT is to modify the Coulomb
interaction: $V(q)=2\pi F(q)/q$, where the formfactor $F$ is
parameterized by an effective thickness.\cite{an82}  Any
second-order correction in terms of $r_{\mbox{\tiny C}}$ would
involve this formfactor doubled and therefore, it would be  more
sensitive to the FT effect. Roughly speaking: a 30\% reduction due
to the FT correction to the spin-flip-mode energy (3.16) causes
the corresponding correction to $\Delta_{\mbox{\tiny ED}}$ [see
Eq. (3.20)] and determines a reduction by $\approx 50\%$ in
$\Delta E^{(2)}$ [also by a factor of $\approx (0.7)^{\kappa}$ in
the following $\Delta E^{(\kappa)}$ corrections]. The two curves
$\Delta_{\mbox{\tiny ED}}$ and $\Delta_{\mbox{\tiny
ED}}^{\mbox{\tiny corr}}$ would start thereby at $r_{\mbox{\tiny
C}}=0$ from $\approx 0.7\sqrt{\pi/8}$ and become more sloping. At
any $r_{\mbox{\tiny C}}$ the gap turns out to be smaller because
of the FT effect.

The disorder may govern QHF features critically and even in the
ground state this can lead to  a realignment of spins with respect
to one another. Specifically, the calculation of disorder effects
depends on the model for the random potential. The white noise
potential (arising e.g. due to chargeless point defects available
in the 2D channel) is  considered in Ref. \onlinecite{si00}. A
perceptible change of the charge gap seems to be related to
appearance  of a skyrmion-like structure in the ground state. The
authors\cite{si00} found that the latter occurs at $\nu=1$
starting from some appreciable threshold for the amplitude of the
disorder potential correlator. In the opposite case of long range
potential fluctuations  (mostly determined by charged impurities
situated out of the spacer), the gap should change smoothly with
correlator amplitude and could be estimated as follows. When
chargeless exciton exists, the disorder broadening determines a
finite cut-off value $q_m$ for 2D momenta: $q<q_m$. This momentum
$q_m$ is related to a certain distance $l_B^2q_m$ (in usual
units). At this distance a force of the Coulomb interaction
between quasiparticles that form the exciton, becomes equal to an
external random force appearing due to the disorder potential.
Hence, the real creation gap for free quasiparticles decreases by
a value $E_x(\infty)-E_x(q_m)$, where $E_x(q)$ is an appropriate
exciton energy calculated within the clean limit (c.f. the
analysis in Refs. \onlinecite{dile99,di00}). It is rather
difficult to estimate the gap reduction corresponding to our
specific case, because the energy with $q$ of a
skyrmion-antiskyrmion exciton is unknown. However, for the spin
exciton\cite{le80,by81,ka84} the analogous estimation \cite{di00}
results in a reduction of $\approx 20\%$, if the random force is
caused by distant impurities. (In real 2D structures this force
could be estimated as $\sim 0.1\;$K/nm.)

Thus, the disorder and FT effects also play a role in the gap
reduction at $r_{\mbox{\tiny C}}\sim 1$. However, for the
up-to-date 2D structures they seem to be less important compared
to the basic effect of the LL mixing.

$\;$

\centerline{\bf ACKNOWLEDGMENTS}

$\;$

Useful discussions with A. M. Finkel'stein, S. V. Iordanskii, K. Kikoin,
and Y. Levinson, are gratefully
acknowledged. The author wishes to thanks  for the hospitality
the Department of Condensed Matter
Physics of The Weizmann Institute of Science (Rehovot), where the
significant
part of this work was carried out. The work was supported by the MINERVA
Foundation and by the Russian Fund for Basic Research.

$\;$

\centerline{\bf APPENDIX I: MATRIX ELEMENTS}

$\;$

The commutation algebra (4.10a-d) for exciton operators (4.6) and (4.8)
allows us with the help of the rule (4.9) to calculate the relevant matrix
elements with relative ease. Using
Eqs. (4.22), (4.4), (4.5), (4.23)-(4.25), (4.29), (4.30),
and in view of the fact that ${\hat Q}_{\alpha\beta}^+\equiv
{\hat {\cal Q}}_{\alpha\beta{\bf 0}}^+$ we find
$$
  \left\langle\alpha\left|{\hat {\cal H}}_{\mbox{\scriptsize {int}}}\right|
  0\right\rangle=\frac{e^2}
  {\varepsilon{}l_B}g_{\alpha}^{(a)}\,, \eqno (\mbox{A}1.1)
$$
where
$$
  g_{\alpha}=
  \frac{V(q_{\alpha})
  (-1)^{n_{a_1}}\left({q_{\alpha}}_-\right)^{n_{a_1}+n_{a_2}}}
  {2\pi\sqrt{n_{a_1}!n_{a_2}!}}e^{-q_{\alpha}^2/2}
  \delta_{\sigma_{a_1},\frac{1}{2}}
  \delta_{\sigma_{a_2},\frac{1}{2}}\,;                      \eqno (\mbox{A}1.2)
$$
and
$$
  \left\langle\alpha\left|\left[{\hat {\cal H}}_{\mbox{\scriptsize {int}}},
  {\hat Q}_{0\overline{1}}^+\right]\right|
  0\right\rangle=\frac{e^2}{\varepsilon{}l_B{N_{\phi}}^{1/2}}
  f_{\alpha}^{(a)}\,,                                    \eqno (\mbox{A}1.3)
$$
where
$$
\begin{array}{lcl}
  \!\!f_{\alpha}=\frac{\displaystyle V(q_{\alpha})
  (-1)^{n_{a_2}}\left({q_{\alpha}}_-\right)^{n_{a_1}+n_{a_2}-1}}
  {\displaystyle 2\pi\sqrt{n_{a_1}!n_{a_2}!}}e^{-q_{\alpha}^2/2}{}\quad{}\qquad\\
  \vphantom{0000}\\
  \!\!{\displaystyle \times}
  \left\{\begin{array}{ccr}0,&\mbox{if}&\sigma_{a_1}+\sigma_{a_2}\not=0,\\
   L_1^{n_{a_1}-1}(q_{\alpha}^2/2),&\mbox{if}&\sigma_{a_1}=
                             -\frac{1}{2},\:\:\sigma_{a_2}=\frac{1}{2},\\
   -L_1^{n_{a_2}-1}(q_{\alpha}^2/2),&\mbox{if}&\sigma_{a_1}=
                             \frac{1}{2},\:\:\sigma_{a_2}=-\frac{1}{2}.\\
  \end{array}{}\;\; (\mbox{A}1.4)
  \right.\\
  \end{array}                             
$$

In the case of the state $|\beta\rangle$ corresponding to $\delta S_z=0$
(i.e. $\sigma_{b_1}=\sigma_{b_2}=1/2$) and in view of Eqs. (3.12),
(4.24), (4.34) and (4.35) we obtain
$$
  \left\langle\alpha\left|\left[{\hat {\cal W}},{\hat P}_{\beta}^+
  \right]\right|0\right\rangle=d_{\alpha\beta}-\frac{1}{N_{\phi}}
  \sum_{\gamma}{\cal F}_{\alpha\gamma}d_{\gamma\beta}\equiv
   d_{\alpha\beta}^{(a)}\,,                       \eqno (\mbox{A}1.5)
$$
where
$$
  d_{\alpha\beta}=\Omega^+_-\left(\sqrt{n_{b_1}+1}\delta_{\alpha,\beta_{1+}}+
  \sqrt{n_{b_2}+1}\delta_{\alpha,\beta_{2+}}\right)
$$
$$
  +\Omega^+_+\left(\sqrt{n_{b_1}}\delta_{\alpha,\beta_{1-}}+
  \sqrt{n_{b_2}}\delta_{\alpha,\beta_{2-}}\right),  \eqno (\mbox{A}1.6)
$$
and indexes $\beta_{1\pm}$ and $\beta_{2\pm}$ designate the states
$$
  \!\!\left|\beta_{1\pm}\right\rangle=\left|\overline{n_{b_1}\pm 1},n_{b_2},
  {\bf q}_{\beta}\right\rangle,\;\,
  \left|\beta_{2\pm}\right\rangle=\left|n_{b_1},
  \overline{n_{b_2}\pm 1},{\bf q}_{\beta}\right\rangle.
$$
[See definitions (4.35), (4.25), and (3.9).] In case $n_{a_1}\!\!+\!n_{a_2}=
n_{b_1}\!\!+\!n_{b_2}$, we find also that for any function
$v_{\alpha}$,
$$
  \sum_{\gamma}v_{\gamma}d^*_{\gamma\alpha}d_{\gamma\beta}=
  \delta_{\alpha,\beta}\left[
  \Omega^+_-\Omega^-_+v_{\beta_{1+}}\left(n_{b_1}+n_{b_2}+2\right)\right. 
$$
$$
  \left.+\Omega^-_-\Omega^+_+v_{\beta_{1-}}\left(n_{b_1}+n_{b_2}\right)\right].
                                                    \eqno (\mbox{A}1.7)
$$

Finally,
$$
  \left\langle\mbox{SF}\left|{\hat {\cal W}}\right|0\right\rangle=
   N_{\phi}^{1/2}\,.
                                                     \eqno (\mbox{A}1.8)
$$

$\;$

\centerline{\bf APPENDIX II: THE COINCIDENCE} 
\centerline{OF THE \mbox{\large {\boldmath
 $r_{\mbox{\tiny C}}\!\to\!0$}} RESULTS at \mbox{\large{\boldmath $\nu=1$}}}

$\;$

We investigate the origin of the coincidence of our results, which are exact
at $r_{\mbox{\tiny C}}\!\to\!0$, with the results obtained: (i) within
the HF and WFP approximations;\cite{by96}
(ii) within the HF approximation\cite{io99,io00}.
In any case the gap has to be proportional to $e^2/\varepsilon l_B$. However,
generally, the specific factor should be different in these three approaches.

In the work\cite{by96}, where only a single LL is considered, the
corresponding factor is determined by the $q^2$ term of the expansion of
${\cal E}_{\mbox{\scriptsize{sw}}}(q)\approx
 q^2{\cal E}_{\mbox{\scriptsize{sw}}}''(0)/2$ at a small
wave vector $q$.
Here
${\cal E}_{\mbox{\scriptsize{sw}}}(q)$ is the energy of spin
exciton\cite{ka84,by81}. [This value is equal to ${\cal E}_{0\overline{0}}(q)$
in our notations, see (3.9b) and (4.15)]. The creation gap in the work of
Ref. \onlinecite{by96} turns out to be equal
to the inverse spin-exciton mass $M^{-1}_{\mbox{\scriptsize{sw}}}=
{\cal E}_{\mbox{\scriptsize{sw}}}''(0)$.

In the work of Ref. \onlinecite{io99} both $n=0$ and $n=1$ LL's
are used for the presentation of the bare one-electron GF. In so
doing, the external HF field for this GF has been taken into
account. Therefore, the denominators of the GF contain the
energies of one electron placed at the $n=1$ LL in its spin ``up''
and spin ``down'' states. The result for the skyrmion-creation gap
turns out to be proportional to the difference of these $n=1$
energies. At the same time, the electron energies can be measured
from the energy of a distant hole at the $n=0$ level. Then, the
required difference corresponds to the extreme case (i.e. to  the
$q\!\to\!\infty$ limit) of the difference of the corresponding
exciton energies:\cite{ka84,lo93} $E_1=\left.\left[ {\cal
E}_{0\overline{1}}(q)-{\cal
E}_{01}(q)\right]\right|_{q\!\to\!\infty}$. [The energies ${\cal
E}_{01}(q)$ and ${\cal E}_{0\overline{1}}(q)$ may be found
directly from Eq. (4.15).]

We remind that our $r_{\mbox{\tiny C}}\!\to\!0$ result for the gap
is ${\cal E}^{\mbox{\tiny{SF}}}_{01}\equiv{\cal E}_{0\overline{1}}(0)$.
Thus, generally, these three different approaches should lead to
the different results. However, due to specific features of the QHF
studied, all these three values actually are
equal to each other:
$$
  M^{-1}_{\mbox{\scriptsize{sw}}}=E_1=
  {\cal E}^{\mbox{\tiny{SF}}}_{01}\equiv
  \frac{e^2}{\varepsilon l_B}
  \int \frac{d^2{\bf q}}{(2\pi)^2}\frac{q^2V(q)}{2}e^{-q^2/2}\,.  
      \eqno (\mbox{A}2.1)
$$
In particular, the coincidence of ${\cal E}^{\mbox{\tiny{SF}}}_{01}$ and $E_1$
is the result of the ``accidental'' equality of the $q=0$ and $q=\infty$
exchange energies for the 01-magnetoplasmon.
Therefore, the Eqs. (\mbox{A}2.1) appear to be nothing more than an
coincidence peculiar to the system studied.

If we study a single skyrmion or antiskyrmion, then we see that their energies
(3.19) are determined also by an additional correction proportional to
$q_{\mbox{\tiny T}}$ (where $q_{\mbox{\tiny T}}=\pm 1$). In fact, this correction
in the present work as well as in Refs. \onlinecite{io99,io00} is determined by
the rotation-matrix feature (3.18) and by the renormalization rule (3.4).
Therefore, under the coincidence condition (A2.1), we arrive again
at the identical results.
The approach of Ref. \onlinecite{by96} seems indirectly to contain also certain
features
analogous to (3.4) and (3.18). Also it results thereby in the same energies
of an isolated skyrmion or antiskyrmion.

Finally, it should be noted that for the filling $\nu \ge 3$ our
result (3.25) differs from the result of  Ref. \onlinecite{by96}.
This fact reflects the role of low lying LL's which participate in
the skyrmion formation. Nevertheless, the $\nu \ge 3$ skyrmion
creation gap, just as in the approach adopted in Ref.
\onlinecite{by96}, turns out to be lower than the corresponding
quasiparticle gap. (In the work of Ref. \onlinecite{io99} only the
$\nu=1$ case was studied.)

$\;$

$\;$

$\;$

$\;$


\begin{references}


\bibitem{so93}
{C. L. Sondhi, A. Karlhede, S. A. Kivelson, and E. H. Rezayi},
Phys. Rev. B {\bf 47},  16419  (1993).

\bibitem{mo95}
{K. Moon, H. Mori, Kun Yang, S. M. Girvin, and
A. H. MacDonald}, Phys. Rev. B {\bf 51},  5138  (1995).

\bibitem{kr95}
{B. Kr\'alik, A. M. Rappe, and S. G. Louie},  Phys. Rev. B {\bf 52},
11626  (1995).

\bibitem{fe94}
{H. A. Fertig, L. Brey, R. C\^ot\'e, and A. H. MacDonald},
Phys. Rev. B {\bf 50},  11018  (1994).

\bibitem{by96}
{Yu. A. Bychkov, T. Maniv, and I. D. Vagner}, Phys. Rev. B {\bf 53},
10148  (1996).

\bibitem{fe97}
{H. A. Fertig, L. Brey, R. C\^ot\'e, and A. H. MacDonald, A. Karlhede,
and S. L. Sondhi}, Phys. Rev. B {\bf 55},  10671  (1997).

\bibitem{io99}
{S. V. Iordanskii, S. G. Plyasunov, and V. I. Falko},
Zh. \'Eksp. Teor. Fiz. {\bf 115}, 716 (1999) [JETP {\bf 88}, 392 (1999)].

\bibitem{io00}
{S. V. Iordanskii and A. Kashuba}, private communication.

\bibitem{ba95}
{S. E. Barret, G. Dabbagh, L. N. Pfeifer, K. W. West, and
R. Tycko}, Phys. Rev. Lett. {\bf 74},  5112  (1995).

\bibitem{ku97}
{I. Kukushkin, K. v. Klitzing, and K. Eberl}, Phys. Rev. B {\bf 60}, 2554
  (1999).

\bibitem{ba96}
{V. Bayot, E. Grivei, S. Melinte, M. B. Santos, and M. Shayegan},
 Phys. Rev. Lett. {\bf 76}, 4584  (1996)

\bibitem{sc95}
{A. Schmeller, J. P. Eisenstein, L. N. Pfeiffer, and K. W. West},
 Phys. Rev. Lett. {\bf 75}, 4290  (1995).

\bibitem{ma96}
{D. K. Maude, M. Potemski, J. C. Portal, M. Heinini, L. Eaves, G. Hill,
and M. A. Pate},  Phys. Rev. Lett. {\bf 77}, 4604  (1996).

\bibitem{le98}
{D. R. Leadley, R. J. Nicholas, D. K. Maude, A. N. Utjuzh,
J. C. Portal, J. J. Harris,
and C. T. Foxon}, Semicond. Sci. Technol. {\bf 13} 671 (1998).

\bibitem{me99}
{S. Melinte, E. Grivei, V. Bayot, and M. Shayegan},
 Phys. Rev. Lett. {\bf 82}, 2764  (1999).

\bibitem{sh00}
{S. P. Shukla, M. Shayegan, S. R. Parihar, S. A. Lion, N. R. Cooper, and A. A.
Kiselev}, Phys. Rev. B {\bf 61}, 4469 (2000).

\bibitem{be75}
{A. A. Belavin and A. M. Polyakov}, Pis'ma Zh. \'Eksp. Teor. Fiz.
{\bf 22}, 503 (1975) [JETP Lett. {\bf 22} 245 (1975)].

\bibitem{ra89}
{R. Rajaraman}, {\it Solitons and Instantons}
(North-Holland, Amsterdam, 1989).

\bibitem{dz83-84}
{A. B. Dzyubenko and Yu. E. Lozovik}, Fiz. Tverd. Tela (Leningrad) {\bf 25},
1519 (1983) [Sov. Phys. Solid State {\bf 25}, 874 (1983)];
Fiz. Tverd. Tela (Leningrad) {\bf 26},
1540 (1983) [Sov. Phys. Solid State {\bf 26}, 938 (1984)].

\bibitem{dz91}
{A. B. Dzyubenko and  Yu. E. Lozovik}, J. Phys. A {\bf 24}, 415 (1991).

\bibitem{by87}
{Yu. A. Bychkov and S. V. Iordanskii},  Fiz. Tverd. Tela (Leningrad) {\bf 29},
2442 (1987) [Sov. Phys. Solid State {\bf 29}, 1405 (1987)].

\bibitem{di96}
{S. Dikman and S. V. Iordanskii}, Zh. \'Eksp. Teor. Fiz. {\bf 110}, 238 (1996)
[JETP {\bf 83}, 128 (1996)].

\bibitem{dile99}
{S. Dickmann and Y. Levinson}, Phys. Rev. B {\bf60} 7760  (1999).

\bibitem{di99}
{S. Dikman and S. V. Iordanskii},  Pis'ma v Zh.
Eksp. Teor. Fiz. {\bf 70}, 531 (1999)
[JETP Lett. {\bf 70}, 543  (1999)].

\bibitem{diE99}
{S. Dickmann and Y. Levinson}, Physica E {\bf 5}, 153 (1999).

\bibitem{di00}
{S. Dickmann}, Phys. Rev. B {\bf 61}, 5461 (2000).

\bibitem{ll91}
{L. D. Landau and E. M. Lifschitz}, {\it Quantum Mechanics}
(Butterworth-Heinemann, Oxford, 1991).

\bibitem{foot1} {The gradient expansion is the expansion in terms of the
small parameter ${}l_B/R$, in which $R$ is a skyrmion radius. The latter is
determined by the ratio of the Zeeman energy to the Coulomb
energy $e^2/\varepsilon{}l_B$.
In our study, the Zeeman energy does not appear. We obtain the results of
the skyrmion energy to the zero order of ${}l_B/R$.}

\bibitem{ka84}
{C. Kallin and B. I. Halperin}, Phys. Rev. B {\bf 30},  5655  (1984).

\bibitem{foot0}
{According to Larmor's theorem there are also pure spin excitations
(having the Zeeman gap) which do not change the space wave function of the
ground state, and therefore they do not influence the activation gap.}

\bibitem{ko61}
{W. Kohn}, Phys. Rev., {\bf 123}, 1242 (1961).

\bibitem{hu00}
{B. Huckestein and M. Backhaus}, cond-mat/0004174 (2000),
to appear in {\it Advances in Solid State Physics Vol. 42},
B. Kramer (ed.), Springer-Verlag, Heidelberg (2002).

\bibitem{chki99}
{S. Chakravarty, S. Kivelson, C. Nayak, and K. Voelker},
Phil. Mag. B {\bf 79}, 859 (1999).

\bibitem{mi00}
{I. Mihalek and H. A. Fertig}, Phys. Rev. B {\bf 62}, 13573 (2000).

\bibitem{foot2}
{For the third Eulerian angle $\zeta$ we have set $\zeta=\varphi$ in
accordance
with Ref. \onlinecite{io99}. Final
results do not depend on $\zeta$, but this substitution enables us to avoid
some non-physical singularities of intermediate results}.

\bibitem{foot3}
{It is noted that the state $|0\rangle$ is degenerate, since any state
$\left(\sigma_{-}\right)^m|0\rangle$ has the same energy to the zero order of
${\hat V}_{\Omega}$. In the first and second approximations, in terms of
${\hat V}_{\Omega}$, we arrive at  a quantum tunneling from $|0\rangle$ to
$\sigma_{-}|0\rangle$. However, this results only in the correction of the
order of $\omega_c\sqrt{N_{\phi}}{}l_B^2\nabla^2$ to the energy that is by 
the factor
$\sim 1/\sqrt{N_{\phi}}$ smaller than the studied result relevant for the
$G_i$ region}.

\bibitem{lo93}
{J. P. Longo and C. Kallin}, Phys. Rev. B {\bf 47},  4429  (1993).

\bibitem{pi92}
{A. Pinczuk, B. S. Dennis, D. Heiman, C. Kallin, L. Brey, C. Tejedor,
S. Schmitt-Rink,
L. N. Pfeiffer, and K. W. West},
Phys. Rev. Lett. {\bf 68},  3623  (1992).

\bibitem{by83}
{Yu. A. Bychkov, and E. I. Rashba}, Zh. Eksp. Teor. Fiz. {\bf 85}, 1826 (1983)
[Sov. Phys. JETP {\bf 58},  1062  (1983)].

\bibitem{foot4}
We can compare these results with the
$r_{\mbox{\tiny C}}^2$ corrections to the energies of the simple $S_z=-1/2$
excitations (quasielectrons and quasiholes). In this case the studied
states are $b^+_p|0\rangle$ and $a_p|0\rangle$, where the sublevel
indexes are $a=(0,+1/2)$ and $b=(0,-1/2)$. The results of the
second-order perturbation theory are obtained with the use of the operator
(4.20). The energy corrections measured from the ground state (which must
be calculated also to the same order) can be found  analytically by
carrying out the corresponding summations:
$E^{(2)}_e=-\frac{\pi^2}{24}-\frac{1}{4}(\ln{2})^2=
-0.531...$ and  $E^{(2)}_h=-\frac{\pi^2}{24}+\frac{3}{4}(\ln{2})^2=-0.0509...$
(in units of $2\,$Ry$=r_{\mbox{\tiny C}}^2\omega_c$).
These values differ slightly from the numerical results presented in Ref. 1.

\bibitem{an82} T. Ando, A. B. Fowler, and F. Stern, Rev. Mod. Phys. {\bf 54},
437 (1982).

\bibitem{si00}
{J. Sinova, A. H. MacDonald, and S. M. Girvin}, Phys. Rev. B {\bf
62}, 13579 (2000).

\bibitem{le80}
{I. V. Lerner and Yu. E. Lozovik}, Zh. Eksp. Teor. Fiz. {\bf 78},
1167 (1980) [Sov. Phys. JETP {\bf 51}, 588 (1980)].

\bibitem{by81}
{Yu. A. Bychkov, S. V. Iordanskii, and G. M. \'Eliashberg},  Pis'ma Zh.
  Eksp. Teor. Fiz. {\bf 33}, 152 (1981)
[Sov. Phys. JETP Lett. {\bf 33}, 143  (1981)].


\end{references}
\end{document}